\documentclass[letterpaper,superscriptaddress,english,aps,prb,twocolumn,floatfix, showpacs, amsfonts, amssymb]{revtex4}
\usepackage{epsfig, graphicx,psfrag,amsmath,amssymb,float}
\usepackage[T1]{fontenc}
\usepackage[latin9]{inputenc}
\usepackage{amsmath}
\usepackage{amssymb}
\usepackage{bbold}
\usepackage{amscd}
\usepackage{bm}
\usepackage{psfrag}
\usepackage{color}
\usepackage{babel}

\begin{document}
\title{Hubbard-model description of the high-energy spin-weight distribution in La$_2$CuO$_4$}

\author{J. M. P. Carmelo}
\affiliation{Center of Physics, University of Minho, Campus Gualtar, P-4710-057 Braga, Portugal}
\affiliation{Institut f\"ur Theoretische Physik III, Universit\"at Stuttgart, D-70550 Stuttgart, Germany}

\author{M. A. N. Ara\'ujo}
\affiliation{Departamento de F\'{\i}sica, Universidade de \'Evora, P-7000-671, \'Evora, Portugal}
\affiliation{CFIF, Instituto Superior T\'ecnico, Universidade T\'ecnica de Lisboa,  Av. Rovisco Pais, 1049-001 Lisboa, Portugal}

\author{S. R. White}
\affiliation{Department of Physics and Astronomy, University of California, Irvine, CA 92617, USA}

\author{M. J. Sampaio}
\affiliation{Center of Physics, University of Minho, Campus Gualtar, P-4710-057 Braga, Portugal}

\date{Published 20 August 2012}

\begin{abstract}
The spectral-weight distribution in recent neutron scattering experiments on the parent compound La$_2$CuO$_4$ (LCO), 
which are limited in energy range to about 450\,meV, is studied in the framework of the Hubbard model on the square lattice
with effective nearest-neighbor transfer integral $t$ and on-site repulsion $U$.
Our study combines a number of numerical and theoretical approaches, including, in addition to standard
treatments, density matrix renormalization group calculations for Hubbard cylinders and a
suitable spinon approach for the spin excitations. The latter spin-$1/2$ spinons are the spins of the
rotated electrons that singly occupy sites. These rotated electrons are mapped from the electrons
by a uniquely defined unitary transformation, in which rotated-electron single and double
occupancy are good quantum numbers for finite interaction values.
Our results confirm that the $U/8t$ magnitude suitable to LCO corresponds to intermediate $U$ values
smaller than the bandwidth $8t$, which we estimate to be $8t \approx 2.36$ eV for $U/8t\approx 0.76$.
This confirms the unsuitability of the conventional linear spin-wave theory. 
Our theoretical studies provide evidence for the occurrence of ground-state $d$-wave spinon pairing in the half-filled Hubbard 
model on the square lattice. This pairing applies only to the rotated-electron spin degrees of freedom, but it could
play a role in a possible electron $d$-wave pairing formation upon hole doping. 
We find that the higher-energy spin spectral weight extends to about $566$ meV and is 
located at and near the momentum $[\pi,\pi]$. 
The continuum weight energy-integrated intensity vanishes or is extremely small at momentum $[\pi,0]$. 
This behavior of this
intensity is consistent with that of the spin waves observed in recent high-energy neutron scattering
experiments, which are damped at the momentum $[\pi,0]$. We suggest that future LCO neutron 
scattering experiments scan the energies between $450$ meV and $566$ meV and momenta around $[\pi,\pi]$.
\end{abstract}


\maketitle

\section{Introduction}
\label{Introduction}

The development of a better understanding of quantum magnetism is important
for improving our understanding of the high-temperature cuprate superconductors. 
Indeed, the parent compounds of the cuprates are insulating antiferromagnets,
and these less complicated undoped systems
can provide valuable information on which model Hamiltonians quantitatively describe
the cuprates. Improved determination of the model Hamiltonians is essential because of
the many nearby competing phases in the doped systems, easily affected by small parameters, 
which can now be seen because of 
continued improvements in numerical simulations \cite{WSstripe}.

The spectral-weight distribution in recent neutron scattering experiments on the 
parent compound La$_2$CuO$_4$ (LCO), which are limited in energy range to about 450\,meV,
raise new interesting questions \cite{headings2010}. In LCO, antiferromagnetic order occurs with 
a commensurate wave vector $[\pi,\pi]$, where $[\pi,\pi]$ is observed to remain commensurate 
for a finite level of doping. A $[\pi,\pi]$ Goldstone mode was predicted by a spin-bag model \cite{SWZ}.
A decade ago the neutron scattering experiments on LCO of Coldea {\it et al.} \cite{LCO-2001} first showed 
sufficient details of the spin-wave spectrum to demonstrate
that a simple nearest-neighbor Heisenberg model must be supplemented by a number of additional
terms, including ring exchanges. 
These terms arise naturally out of a single band Hubbard model with finite $U/t$,
and several detailed studies showed that the spin-wave data in the available energy window 
could be successfully described by the Hubbard model
using a somewhat smaller value of $U/t \sim 6-8$ than originally thought
appropriate \cite{LCO-2001,peres2002,lorenzana2005,companion}. 
(For the effective Coulomb repulsion $U$ in units of the bandwidth, $8t$, this refers to intermediate values, 
$U/8t \sim 0.75-1$.) 

Part of the spin spectral weight reported in  Ref. \cite{LCO-2001} was deduced to be outside 
the energy window. The recent improved neutron scattering experiments of Ref. \cite{headings2010},
with a wider energy window of about 450\,meV, have raised a number of questions. 
Surprisingly, these studies revealed that the high-energy spin waves are 
strongly damped near momentum $[\pi,0]$ and merge into a momentum-dependent continuum.
These results led the authors of Ref. \cite{headings2010} to conclude that
``the ground state of La$_2$CuO$_4$ contains additional correlations not captured by 
the N\'eel-SWT [spin-wave theory] picture''.

This raises the important question of whether the more detailed results can still be
described in terms of a simple Hubbard model. We show that the Hubbard model {\it does} describe 
the new neutron scattering results. Our results confirm that the $U/t$ value suitable to LCO is in the range 
$U/t\in (6,8)$. Inclusion of second- and third-neighbor hopping parameters, $t'$ and $t''$, into
the Hubbard Hamiltonian lead to an interaction strength $U/t\approx 8$. Specifically,
the studies of Ref. \cite{Tremblay-09} have considered that the best fits to the ensemble
of LCO inelastic neutron scattering points from Ref. \cite{LCO-2001} are reached
for $U/t\approx 7.9$ if one includes four independent parameters, $t$, $t'$, $t''$, $U$, 
and for $U/t\approx 7.1$ if one includes only $t$ and $U$. Furthermore, the results of
Refs. \cite{Tremblay-09,Miguel-03} reveal that as far as the LCO inelastic neutron scattering is concerned the 
inclusion of $t'$ and $t''$ does not lead to a better quantitative fit.
Accordingly, the studies of this paper consider the half-filled Hubbard model on the
square lattice with only two independent effective parameters, $t$ and $U$.

Our study uses a combination of a number of numerical and theoretical approaches, including, in addition to standard
treatments, density matrix renormalization group (DMRG) calculations for 
Hubbard cylinders \cite{Steve-92,DMRG-SL-2011,DMRG-tJ-2011}
and, since conventional linear spin-wave theory is unsuitable,
a spinon operator approach for the spin excitations \cite{companion,companion0}.
The spinon operator approach is suitable for LCO's intermediate $U/t$ values, and
corresponds to a particular case of a general operator representation
that profits from the recently found model's extended global symmetry \cite{bipartite}.

An exact result valid for the Hubbard model on any bipartite lattice is that for onsite interaction $U\neq 0$ it has two global 
$SU(2)$ symmetries \cite{HL,Lieb89}, which refer to a global $SO(4)=[SU(2)\otimes SU(2)]/Z_2$ symmetry \cite{Yang89,Zhang}.
A recent study of the problem by one of us and collaborators reported in Ref. \cite{bipartite}, reveals that an exact extra global $c$ hidden 
$U(1)$ symmetry emerges for $U\neq 0$, in addition to the $SO(4)$ symmetry. Specifically, the Hubbard model on a bipartite lattice, such as the 
present square lattice, has a global $[SU(2)\otimes SU(2)\otimes U(1)]/Z_2^2=[SO(4)\otimes U(1)]/Z_2=SO(3)\otimes SO(3)\otimes U(1)$ 
symmetry. The index $c$ in the designation {\it $c$ hidden symmetry} is intended to distinguish it from the $\eta$-spin $U(1)$ symmetry
and spin $U(1)$ symmetry in the corresponding two model's $SU(2)$ symmetries. The index $c$ also labels the $c$ fermions,
whose occupancy configurations generate the representations of the global $c$ hidden $U(1)$ symmetry algebra. That the
latter symmetry is {\it hidden} follows from the fact that except in the $U/t\rightarrow\infty$ limit,
 its generator does not commute with
the electron - rotated-electron unitary operator. As a result, for finite $U/t$ values its expression in terms of electron creation and annihilation
operators has an infinite number of terms. (The generators of the $\eta$-spin and spin $SU(2)$ symmetries commute
with that unitary operator.)

The origin of the extended global symmetry is a local gauge $SU(2)\otimes SU(2)\otimes U(1)$ symmetry of the model 
Hamiltonian electron-interaction term first identified in Ref. \cite{U(1)-NL}. 
That local symmetry becomes for finite $U$ and $t$ a group of permissible unitary transformations. 
The corresponding local $U(1)$ canonical transformation is not the ordinary $U(1)$ gauge subgroup of electromagnetism. 
It is rather a ``nonlinear" transformation \cite{U(1)-NL}.

For very large $U/t$ values the Hubbard model may be mapped onto
a spin-only problem whose spins are those of the electrons that
singly occupy sites. However, for intermediate $U/t$ values this mapping
generates many complicated terms in the Hamiltonian, when written in terms of electron
creation and annihilation operators. Here we address that problem by expressing
the Hamiltonian in terms of the rotated-electron operators, which  naturally emerge from 
the generators of the model's symmetries. 

In contrast to electrons, for rotated electrons single and double occupancy are
good quantum numbers for $U/t>0$.  For large $U/t$ values
electrons and rotated electrons are the same objects. 
Apparently, the Hamiltonian $t/U$ expansion is formally similar in terms of electron
and rotated-electron operators. However that is only so for very large $U/t$ values.
For instance, there are well-defined $t^2/U$ and $t^4/U^3$ terms in the Hamiltonian
expression in terms of rotated-electron operators,
which are identical in form for very large $U/t$ to the corresponding terms using  electron operators.
However, for intermediate $U/t\in (6,8)$, if one expresses the former $t^2/U$ and $t^4/U^3$ terms in
electron creation and annihilation operators, one finds many complicated higher order
$t^j/U^{j-1}$ terms where for the half-filled case $j$ are even integers,
$j=2,4,6,...$ and $j=4,6,8,...$, respectively. Hence the first few terms of
the Hamiltonian expression in terms of rotated-electron operators describe many higher-order 
electron processes. For moderate $U/t$ the rotated-electron operators also
generate a much simpler form for the energy eigenstates as well as for
complicated processes involving a large number of electrons.
Our spin-$1/2$ spinons correspond to the spin-$1/2$ spins of the rotated electrons that singly
occupy sites, so that they are well defined for $U/t>0$. 

When one decreases $U/t$ to the intermediate $U/t$ values suitable for LCO, 
the above mentioned Hamiltonian terms become increasingly important and generate higher-order spinon 
processes. Fortunately, those are simpler than the corresponding electron processes. Indeed,
the use of our operational representation renders the intermediate $U/t$ quantum problem in terms 
of rotated electrons similar to the corresponding large-$U/t$ quantum problem in terms 
of electrons. The effect of decreasing $U/t$ is mostly an increase of the energy bandwidth of
an effective band associated with the spinon occupancy configurations. The intermediate $U/t$ rotated-electron
processes may be associated with exchange constants 
describing rotated-electron motion touching progressively larger number of sites. 
Within our Hubbard model's representation such Hamiltonian terms emerge
naturally upon decreasing the magnitude of $U/t$.

Our theoretical studies provide evidence of the occurrence of 
ground-state $d$-wave spinon pairing in the half-filled Hubbard model on the square lattice.
One of the few exact theorems that applies to the half-filled Hubbard model on a
bipartite lattice with a finite number of sites and thus on a square lattice is that its ground state is a spin-singlet
state \cite{Lieb89}. Within our spinon representation the ground-state spin-singlet $N$-spinon configuration 
corresponds to $N/2$ independent spin-neutral two-spinon configurations. Under the spin-triplet excitation one of the $N/2$ 
spin-singlet spinon pairs is broken. Quantitative agreement with the spin-wave spectrum obtained from our standard
many-particle diagrammatic analysis is reached provided that the broken
spinon pair has $d$-wave pairing in the initial ground state. Such a pairing refers only 
to the rotated-electrons spin degrees of freedom. However, it could play a role in a possible 
$d$-wave electron pairing formation upon hole doping. 

Applying our approach to the new LCO high-energy neutron scattering 
reported in Ref. \cite{headings2010}, we find that at momentum $[\pi,0]$ 
the continuum weight energy-integrated intensity vanishes or is extremely small. Furthermore, 
we find that beyond 450\,meV, the spectral weight is mostly located around momentum $[\pi,\pi]$ 
and extends to about 566\,meV, suggesting directions for future experiments.

The paper is organized as follows: In Sec. \ref{model} the Hubbard model on the square
lattice and the basic quantities of our study are introduced. The description of the model's antiferromagnetic 
long-range order is addressed in Sec. \ref{AF-order}. Sec.
\ref{spin-spectrum-cohe} presents a random-phase-approximation (RPA) study of
the model's coherent spin-wave spectrum and intensity. Quantitative agreement with
that observed in neutron-sattering experiments is used to find the $U$ and $t$ values
suitable to LCO. The rotated-electron description emerging from the Hubbard model 
on the square lattice with extended global symmetry is introduced in Sec. \ref{general-rep}.
This is the only section where general electronic densities and spin densities
are considered. The goal of this more general analysis is the introduction
of a spinon representation suitable to the LCO intermediate values of $U/t$.
In Sec. \ref{spin-spectrum-incoh} this spinon representation is used
in the study of the general spin-triplet spectrum of the half-filled Hubbard
model on the square lattice, which includes both the
spin waves and the incoherent spin-weight continuum distribution. The
comparison of the predicted spectral weights with those observed in the LCO
high-energy neutron scattering is the goal of Sec. \ref{LCOspin-spectrum}.
Finally, Sec. \ref{concluding} contains the concluding remarks.

\section{The model and the basic quantities of our study}
\label{model}

Most of our results refer to half-filling, so that the number of lattice sites, $N_a$, equals the number of
electrons $N$. The exception is the general analysis reported in Sec. \ref{general-rep}, 
which considers arbitrary values of the electronic density $n=N/N_a$. 
The Hubbard model on a square lattice with $N_a\gg 1$ sites and periodic boundary conditions reads,
\begin{eqnarray}
{\hat{H}} & = & t\,\hat{T}+U\,{\hat{V}}_D \, ,
\nonumber \\
\hat{T} & = & = -\sum_{\langle j,j'\rangle}\sum_{\sigma}(c_{\vec{r}_j,\sigma}^{\dag}\,c_{\vec{r}_{j'},\sigma} + 
c_{\vec{r}_{j'},\sigma}^{\dag}\,c_{\vec{r}_j,\sigma}) \, ,
\nonumber \\
{\hat{V}}_D & = & \sum_{j=1}^{N_a}\left(\hat{n}_{\vec{r}_j,\uparrow} -1/2\right)
\left(\hat{n}_{\vec{r}_j,\downarrow}-1/2\right) \, .
\label{HH}
\end{eqnarray}
Here $\hat{T}$ is the kinetic-energy operator in units of $t$, ${\hat{V}}_D$ is the on-site repulsion interaction
operator in units of $U$, $c_{\vec{r}_j,\sigma}^{\dag}$ and $c_{\vec{r}_{j},\sigma}$ are electron creation and 
annihilation operators with site index $j=1,...,N_a$ and spin $\sigma =\uparrow,\downarrow$, and
$\hat{n}_{{\vec{r}}_j,\sigma} = c_{\vec{r}_j,\sigma}^{\dag} c_{\vec{r}_j,\sigma}$. The on-site repulsion interaction
operator ${\hat{V}}_D$ may alternatively be expressed in terms of the electron double-occupancy operator ${\hat{D}}$ 
or single-occupancy operator ${\hat{Q}}$ given by,
\begin{equation}
{\hat{D}}= ({\hat{N}}-{\hat{Q}})/2 \ ;
\hspace{0.50cm}
{\hat{Q}} = \sum_{j=1}^{N_a}\sum_{\sigma =\uparrow
,\downarrow}\,\hat{n}_{\vec{r}_j,\sigma}\,(1- \hat{n}_{\vec{r}_j,-\sigma}) \, ,
\label{DQ}
\end{equation}
respectively.
The expectation values,
\begin{eqnarray}
d & = & {1\over N_a}\sum_{j=1}^{N_a}\langle GS\vert\hat{n}_{\vec{r}_j,\uparrow}\hat{n}_{\vec{r}_j,\downarrow}\vert GS \rangle \, ,
\nonumber \\
(1-2d)  & = & {1\over N_a}\sum_{j=1}^{N_a}\langle GS\vert(\hat{n}_{\vec{r}_j,\uparrow}-\hat{n}_{\vec{r}_j,\downarrow})^2\vert GS\rangle \, ,
\nonumber \\
m_{AF} & = & {1\over N_a}\sum_{j=1}^{N_a} {1\over 2}\langle GS\vert (-1)^j (\hat{n}_{\vec{r}_j,\uparrow}-\hat{n}_{\vec{r}_j,\downarrow})\vert GS\rangle
\nonumber \\
& \approx  & [1-2\delta S]\,m_{AF}^0 \, ,
\label{d-m}
\end{eqnarray}
play an important role in our study, following the strong evidence that for $U>0$ and 
$N_a\rightarrow\infty$ the model's ground state has antiferromagnetic long-range order \cite{Mano}. 
In the last expression of Eq. (\ref{d-m}) one has that $j$ is an even integer and an odd integer for each of the two sub-lattices,
respectively. Moreover, in that expression $m_{AF}^0$ stands for
a mean-field sub-lattice magnetization that does
not account for the effect of transverse fluctuations while $\delta S$
does account for this effect, its value being estimated below. 
Specifically, $m_{AF}^0$ is the sub-lattice magnetization of the spin-density wave (SDW) 
state obtained in a standard mean-field treatment of the Hubbard interaction 
at zero absolute temperature as given, for instance, in Fig. 3 of 
Ref. \cite{peres2002}.

The on-site spin operators involved in our studies read,
\begin{eqnarray}
{\hat{s}}^{x}_{\vec{r}_j,s} & = & {1\over 2}[{\hat{s}}^{+}_{\vec{r}_j,s} +{\hat{s}}^{-}_{\vec{r}_j,s}] 
\, ; \hspace{0.35cm}
{\hat{s}}^{y}_{\vec{r}_j,s} = {1\over 2i}[{\hat{s}}^{+}_{\vec{r}_j,s} -{\hat{s}}^{-}_{\vec{r}_j,s} ] \, ,
\nonumber \\
{\hat{s}}^{+}_{\vec{r}_j,s} & = & c_{\vec{r}_j,\downarrow}^{\dag}\,c_{\vec{r}_j,\uparrow}
\, ; \hspace{0.35cm}
{\hat{s}}^{-}_{\vec{r}_j,s} = c_{\vec{r}_j,\uparrow}^{\dag}\,c_{\vec{r}_j,\downarrow} \, ,
\nonumber \\
{\hat{s}}^{z}_{\vec{r}_j,s} & = & -{1\over 2}[{\hat{n}}_{\vec{r}_j,\uparrow}-{\hat{n}}_{\vec{r}_j,\downarrow}] \, .
\label{S-j-s}
\end{eqnarray}
The index $s$ in these operators distinguishes them from the corresponding local
operators associated with the $\eta$-spin $SU(2)$ symmetry algebra considered below
in Sec. \ref{general-rep}. 
\begin{figure}[hbt]
\begin{center}
\centerline{\includegraphics[width=6.00cm]{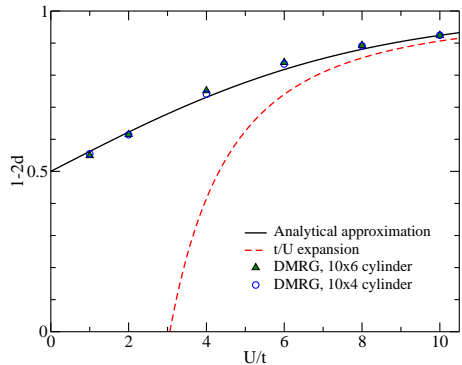}}
\caption{Average single-occupancy: approximate 
expression $[1+\tanh (U/8t)]/2$ valid for $U/t\leq 8$ (solid line),
the limiting $U/t\gg 1$ expression $[1 -c_0 (8t/U)^2]$ (dashed line),
and from DMRG numerical results on two different width cylinders.}
\label{fig1}
\end{center}
\end{figure}

Our study also involves the spin dynamical structure factors,
\begin{eqnarray}
& & S^{\alpha\alpha'} (\vec{k},\omega) = {(g\mu_B)^2\over N_a}\sum_{j,j'=1}^{N_a}e^{-i\vec{k} (\vec{r}_j-\vec{r}_{j'})}
\nonumber \\
& \times &
\int_{-\infty}^{\infty}dt\,e^{i\omega t}
\langle GS\vert {\hat{s}}^{\alpha}_{\vec{r}_j,s} (t) {\hat{s}}^{\alpha'}_{\vec{r}_{j'},s} (0)\vert GS\rangle \, ,
\label{SS}
\end{eqnarray}
where $\alpha =\alpha'=x,y,z$ or $\alpha =-$ and $\alpha' =+$ and
below we consider $g=2$. It is straightforward to show that the sum rules
$[1/N_a]\sum_{\vec{k}}\,S^{\alpha\alpha'} (\vec{k})$ 
where $S^{\alpha\alpha'} (\vec{k})= [1/2\pi]\int_{-\infty}^{\infty}d\omega\,S^{\alpha\alpha'} (\vec{k},\omega)$
involve the average single occupancy $(1-2d)$ and read,
\begin{equation}
{1\over N_a}\sum_{\vec{k}}\,S^{\alpha\alpha'} (\vec{k}) =
{(g\mu_B)^2\over 4}[\delta_{\alpha,\alpha'}+2\delta_{\alpha,-}\delta_{\alpha',+}] (1-2d) \, .
\label{sum-rule-aap}
\end{equation}

In an ideal experiment all components are detected with equal
sensitivity. As discussed for instance in Ref. \cite{lorenzana2005}, in that case 
a transfer of spectral weight from the longitudinal to the transverse part as 
the energy increases is observed. Hence independent of the scattering geometry, the corresponding
effective spin dynamical structure factor satisfies the sum rule,
\begin{equation}
{1\over N_a}\sum_{\vec{k}}{1\over 2\pi}\int_{-\infty}^{\infty}d\omega\,
S^{exp} (\vec{k},\omega) = \mu_B^2\,2(1-2d) \, .
\label{sr-eff}
\end{equation}
That the coefficient involved is $2(1-2d)$ rather than $3(1-2d)$ follows from 
one mode being perpendicular to the plane and thus silent in the experiment \cite{lorenzana2005}. 

\section{The Hubbard model on the square lattice antiferromagnetic long-range order}
\label{AF-order}

For the range $U/t\in (0,8)$, the antiferromagnetic long-range order may be accounted for by a variational 
ground state with a SDW initial trial state, such as for instance a Gutzwiller projected antiferromagnetic
state \cite{Dionys-09},
\begin{equation}
\vert G\rangle =e^{-g\hat{D}}\vert SDW\rangle \, , \hspace{0.15cm} U/t<8 \, ,
\label{G-SDW}
\end{equation}
or the following related state, 
\begin{equation}
\vert GB\rangle =e^{-h\hat{T}/t}e^{-g\hat{D}}\vert SDW\rangle \, , \hspace{0.15cm} U/t<8 \, .
\label{GB-SDW}
\end{equation}
Here $\vert SDW\rangle$ is the ground state of a simple effective mean-field Hamiltonian, such as 
that of Eq. (18) of Ref. \cite{Dionys-09}. For $U/t\gg 1$ this order is as well accounted for 
by a Baeriswyl variational state,
\begin{equation}
\vert B\rangle =e^{-h\hat{T}/t}\vert\infty\rangle \, , \hspace{0.15cm} U/t\gg 1 \, ,
\label{B-SDW}
\end{equation}
where $\vert\infty\rangle$ is the exact $U/t\rightarrow\infty$ ground state \cite{Dionys-09}.
The coefficients $h$ and $g$ multiplying the kinetic-energy and
double-occupancy operators, respectively, in the state expressions given
in Eqs. (\ref{G-SDW})-(\ref{B-SDW}) are variational parameters. 
The above states involve as well a variational gap parameter $\Delta$, which
is expected to tend to zero as the trial state approaches the exact ground
state. Indeed, that variational parameter is an infinitesimal symmetry-breaking
field.

Similarly for the trial state $\vert SDW\rangle$, the relation
\begin{equation}
4[m_{AF}^0]^2 = (1-4d) \, ,
\label{1m-2}
\end{equation}
holds for the states $\vert G\rangle$, $\vert GB\rangle$, and $\vert B\rangle$. However, the 
corresponding function $d=d(U/t)$ is in general state dependent. Inversion of the simple
relation provided in Eq. (\ref{1m-2}) gives,
\begin{equation}
d = {1\over 4}[1-4[m_{AF}^0]^2] \, .
\label{d}
\end{equation}
This is consistent with $d$ not being affected by transverse fluctuations.

The evaluation of the ground-state energy for $\vert G\rangle$ and $\vert GB\rangle$ is for $N_a\gg 1$ 
an involved problem. Here we resort to an approximation, which corresponds to the simplest 
expression of the general form,
\begin{equation}
E/N = T_0\,q_U  + U d \, ; \hspace{0.5cm} T_0 = -{16\over\pi^2}\,t \, ,
\label{E-T0}
\end{equation}
compatible with three requirements. Those are:\\
\vspace{0.05cm}

1) The relation $d = {1\over 4}[1-4[m_{AF}^0]^2]$ provided in Eq. (\ref{d})
must be fulfilled;\\
\vspace{0.05cm}

2) The antiferromagnetic long-range order must occur for the whole $U/t>0$ range;\\
\vspace{0.05cm}

3) The small-$U/t$ expansion of the energy $E/N$ of Eq. (\ref{E-T0}) must lack of a linear
kinetic-energy term in $U$ for $U/t\ll 1$ (except for the term $U d$ corresponding to the
on-site repulsion.)\\
\vspace{0.05cm}

Brinkman and Rice found $q_U=8d (1-2d)$ 
for the original paramagnetic-state Gutzwiller approximation \cite{BR-70}, which is lattice insensitive and thus does not 
account for the square-lattice antiferromagnetic long-range order.  
The simplest modified form of the quantity $q_U$ suitable to a broken-symmetry
ground state such that the above three conditions are met is,
\begin{equation}
q_U = \left({U\over 8t}\right)\,a_{1}^{(+)}\,d\left[{(1-2d)\over  4[m_{AF}^0]^2}- a_2\right] - a_3 \, .
\label{qU}
\end{equation}
Here,
\begin{equation}
a_{1}^{(\pm)} = \pi^2 \pm 4 \, ; \hspace{0.5cm}
a_2 = (1-[\pi^2/2a_1^{(+)}]) \, ,
\label{a1-a2}
\end{equation}
and the coefficient $a_3$ is a function  $a_3 = a_3 (U/t)$ of $U/t$ whose approximate limiting behaviors are,
\begin{eqnarray}
a_3 & = & {a_{1}^{(-)}\over 8}\left[1-\tanh \left({U\over 8t}[(4+a_{1}^{(+)})/a_{1}^{(-)}]\right)\right] 
\, , \hspace{0.15cm} U/t<8 \, ,
\nonumber \\
& = & -c_0[\pi/2]^2\,{8t\over U} \, , \hspace{0.15cm} U/t\gg 1 \, ,
\label{a3}
\end{eqnarray}
where, 
\begin{equation}
c_0 ={1\over 2}\left[{\alpha\over 4} +{1\over 8}\right] = 0.1462 \, .
\label{c0}
\end{equation}
The corresponding estimate $\alpha =0.6696$ is that of the Heisenberg-model studies of Ref. \cite{Mano}.
Moreover, the quantity $4[m_{AF}^0]^2$ on the right-hand side of Eq. (\ref{qU}) behaves as $4[m_{AF}^0]^2=U/8t$ 
for $U/t\ll 1$.

Minimization of the ground-state energy defined by Eqs. (\ref{E-T0})-(\ref{a3}) with respect
to $d$ leads indeed to $d = {1\over 4}[1-4[m_{AF}^0]^2]$. The limiting behaviors of
that energy are,
\begin{eqnarray}
E/N & \approx & T_0 + U d - {1\over 8\pi^2}\,{U^2\over t}
\, , \hspace{0.15cm} U/t\ll 1 \, ,
\nonumber \\
& \approx & -4c_0\,{8t^2\over U} \, , \hspace{0.15cm} U/t\gg 1 \, ,
\label{E-limits}
\end{eqnarray}
We note that the small-$U/t$ second-order coefficient reads $-[1/8\pi^2]\approx -0.0127$,
in agreement with that, $\approx -0.0127$, obtained by second-order perturbation theory \cite{MZ-89}. 
For $U/t\gg 1$ one recovers the known result $E/N =-4c_0[8t^2/U]$ \cite{Mano}, so that our approximation agrees with
the known limiting behaviors.

The quantity $4[m_{AF}^0]^2$ on the right-hand side of Eq. (\ref{qU}) behaves as $4[m_{AF}^0]^2=U/8t$ 
for $U/t\ll 1$. However, its $U/t$ dependence for the range $U/t\in (0,8)$ remains an open problem. Here we
have performed DMRG calculations of $(1-2d)$, which according to the relation provided in Eq. (\ref{d})
is given by $(1-2d) = {1\over 2}(1+ 4[m_{AF}^0]^2)$. Hence its $U/t$ dependence fully determines
that of $4[m_{AF}^0]^2$. The corresponding DMRG results are shown in Fig. 1.
Specifically, two different circumference cylinders were simulated as a function of $U/t$,
with open boundary conditions in $x$ and periodic in $y$, and the double occupancy measured in one
of the middle columns. A maximum of 
$m=4000$ states were kept, with an accuracy of $\sim 10^{-4}$ in $(1-2d)$ for the $10\times4$ system
for the least accurate smaller $U/t$ values,
and about $10^{-3}$ for the $10\times6$ system.
We find that the value of $(1-2d)$ is relatively insensitive to cluster size, and these cluster
sizes are representative of two-dimensional (2D) behavior \cite{Paiva}.

We then find that $4[m_{AF}^0]^2\approx \tanh (U/8t)$ gives for the range $U/t\in (0,8)$ quantitative agreement 
for the $(1-2d)$ dependence on $U/t$ with both our numerical 
DMRG calculations (see Fig. 1) and the numerical results for the states $\vert G\rangle$ and $\vert GB\rangle$
(see Fig. 4 of Ref. \cite{lorenzana2005}). For $U/t\gg 1$ we find the behavior
$4[m_{AF}^0]^2\approx e^{-2c_0\,(8t/U)^2}$ for the state $\vert B\rangle$, so that,
\begin{eqnarray}
(1-2d) & \approx & {1\over 2}\left[1+\tanh\left({U\over 8t}\right)\right] 
\, , \hspace{0.15cm} U/t<8 \, ,
\nonumber \\
& \approx & 1 -c_0 \left({8t\over U}\right)^2 \, , \hspace{0.15cm} U/t\gg 1 \, .
\label{1-2d-limits}
\end{eqnarray}

Furthermore, the states $\vert SDW\rangle$ and $\vert G\rangle$ give a
sub-lattice magnetization $m_{AF}\approx m_{AF}^0={1\over 2}\sqrt{1-4d}$, with
an improved $U/t$ dependence, $d \approx {1\over 4}[1-\tanh (U/8t)]$, for the latter,
as follows from the corresponding $(1-2d)$ expression of Eq. (\ref{1-2d-limits}). For the
state $\vert G\rangle$ the sub-lattice magnetization is then given by,
\begin{equation}
m_{AF}^0 \approx {1\over 2}\sqrt{\tanh\left({U\over 8t}\right)}
\, , \hspace{0.25cm} U/t<8 \, .
\label{M-AF-0}
\end{equation}
On the other hand, we find that the 
states $\vert GB\rangle$ and $\vert B\rangle$ have $m_{AF}^{GB}$ and $m_{AF}^{B}$  
sub-lattice magnetization numerical values very close to those given by the
relation $[1-2\delta S]\,m_{AF}^0$ of Eq. (\ref{d-m}) with, 
\begin{eqnarray}
\delta S & \approx & d \, , \hspace{0.15cm} U/t<8 \, ,
\nonumber \\
& \approx & d+{1\over 2}\left[1-{m_{HAF}\over m_{HAF}^0}\right] \, , \hspace{0.15cm} U/t\gg 1 \, ,
\label{dS-limits}
\end{eqnarray}
respectively. Here $m^0_{HAF}=1/2$ and $m_{HAF} \approx 0.303$ is the Heisenberg-model's sub-lattice magnetization  
magnitude \cite{Mano}, so that $\delta S \approx d+0.197$ in Eq. (\ref{dS-limits}) for $U/t\gg 1$. Hence
one finds,
\begin{eqnarray}
m_{AF}^{GB} & \approx & {1\over 4}\left[1+\tanh\left({U\over 8t}\right)\right]\sqrt{\tanh\left({U\over 8t}\right)}
\, , \hspace{0.10cm} U/t<8 \, ,
\nonumber \\
m_{AF}^{B} & \approx &  \left[0.303 -0.803\times c_0 \left({8t\over U}\right)^2\right] \, , \hspace{0.10cm} U/t\gg 1 \, .
\label{M-AF-limits}
\end{eqnarray}
The magnitudes of the sub-lattice magnetizations $m_{AF}^0$ 
and $m_{AF}^{GB}$ as given in Eqs. (\ref{M-AF-0}) and (\ref{M-AF-limits}) 
for the states $\vert G\rangle$ and $\vert GB\rangle$, respectively,
are provided in Table \ref{table1} for several $U/t$ values. In that table the magnitudes of
a sub-lattice magnetization $m_{AF}^{lower}$ that for $U/t>0$ we define as
$m_{AF}^{lower}=(1-2d)[m_{HAF}/m_{HAF}^0]\,m^0_{AF}$
are also given. Note that for $U/t\gg 1$ the sub-lattice magnetization $m_{AF}^{lower}$
becomes $m_{AF}^{B}$. Probably it is closest to the exact $m_{AF}$, while $m_{AF}^{GB}$ is 
that consistent with our use of the RPA in the ensuing section, to study the spin-wave
spectrum and corresponding intensity.
\begin{table}
\begin{tabular}{|c|c|c|c|c|c|} 
\hline
$U/t$ & $6.1$ & $6.5$ & $8.0$ & $10.0$ \\
\hline
$m^{0}_{AF}$ & $0.401$ & $0.410$ & $0.436$ ($0.43$ \cite{lorenzana2005,Dionys-09}) & $0.461$ ($0.456$ \cite{lorenzana2005}) \\
\hline
$m^{GB}_{AF}$ & $0.329$ & $0.342$ & $0.384$ ($0.39$ \cite{Dionys-09}) & $0.426$ \\
\hline
$m^{lower}_{AF}$ & $0.200$ & $0.207$ & $0.233$ & $0.258$ \\
\hline
\end{tabular}
\caption{The sub-lattice magnetizations as calculated here for several $U/t$ values and
some results from Refs. \cite{lorenzana2005,Dionys-09}.}
\label{table1}
\end{table} 

\section{Coherent spin-wave spectrum and intensity and LCO  $U$ and $t$ values}
\label{spin-spectrum-cohe}

To study the coherent spin-wave weight distribution and spectrum, we have calculated 
the transverse dynamical susceptibility,
\begin{eqnarray}
& & \chi^{-+} (\vec{k},\tau) = {(g\mu_B)^2\over N_a} \sum_{j,j'=1}^{N_a} 
e^{-i\vec{k}\cdot\left(\vec{r}_j-\vec{r}_{j}'\right)} 
\nonumber \\
& \times &
\langle {\hat{s}}^-_{\vec{r}_j,s} (\tau) {\hat {s}}^+_{\vec{r}_{j'},s} (0)\rangle \, ,
\label{Chi-+}
\end{eqnarray}
in the RPA. Here $\tau$ denotes the imaginary time in Matsubara formalism and we shall take the zero temperature limit. 
Because we deal with the antiferromagnetic order (N\'eel state)  it is convenient to define two sub-lattices, $a$ and $b$,  
and redefine the susceptibility as a 2$\times$2 tensor $\tilde\chi_{\mu,\nu}$ where the Greek 
subscripts denote sub-lattice indices,
\begin{eqnarray}
& & \tilde\chi_{\mu,\nu}^{-+} (\vec{k},\tau) = \frac{(g\mu_B)^2}{(N_a/2)} \sum_{j\in \nu,j'\in \mu}e^{-i\vec{k}\cdot\left(\vec{r}_j-\vec{r}_{j}'\right)}
\nonumber\\ 
& \times &  \langle {\hat{s}}^-_{\vec{r}_j,s} (\tau) {\hat{s}}^+_{\vec{r}_{j'},s} (0)\rangle \, .
\label{Chimunu}
\end{eqnarray}
The original  susceptibility in Eq. (\ref{Chi-+}) is then simply related to this tensor as,
\begin{equation}
\chi(\vec k,\tau)= \frac 1 2  \left[   \tilde\chi_{aa}+ \tilde\chi_{bb} + \tilde\chi_{ab}+ \tilde\chi_{ba}    \right] \, .
\end{equation}

We define electron field operators for 
each sub-lattice, $\hat a_{\vec k \sigma}$ and $\hat b_{\vec k \sigma}$, as,
\begin{eqnarray}
\hat c_{\vec r_j \in a, \sigma} &=& \frac{1}{\sqrt{N_a/2}} \sum_{\vec k \in RBZ} e^{i\vec k \cdot\vec r_j} \hat a_{\vec k \sigma} \, , \\
\hat c_{\vec r_j \in b, \sigma} &=& \frac{1}{\sqrt{N_a/2}} \sum_{\vec k \in RBZ} e^{i\vec k \cdot\vec r_j} \hat b_{\vec k \sigma} \, .
\end{eqnarray}
In the momentum summations the reduced Brillouin zone (RBZ) covers only half of the original Brillouin zone (BZ) for the square lattice.
The effective Hamiltonian that describes the SDW phase in mean field theory for the Hubbard interaction 
can be written as,
\begin{equation}
\hat H_{eff} = \sum_{\vec k , \sigma} 
\left( \begin{array}{cc} \hat a_{\vec k \sigma}^\dagger & \hat b_{\vec k \sigma}^\dagger    \end{array}\right)
\left( \begin{array}{cc} \epsilon_\sigma  & f(\vec k) \\ 
                          f(\vec k)  & -\epsilon_\sigma  \end{array}\right)
\left( \begin{array}{c}   \hat a_{\vec k \sigma} \\  \hat b_{\vec k \sigma}    \end{array}\right) \, ,
\label{Heff}
\end{equation}
with,
\begin{eqnarray}
\epsilon_\sigma & = & -U \,{\sigma\over 2}\,m_{AF} \, ,
\nonumber \\
f(\vec k) & = & -2t\left[\cos(k_x) + \cos(k_y)\right] \, .
\label{epsilon-fk}
\end{eqnarray}
Using the effective Hamiltonian given in Eq. (\ref{Heff}) one derives from Eq. (\ref{Chimunu})
a susceptibility tensor $\tilde\chi_{\mu,\nu}^{(0)}(\vec k,\tau)$ at  mean field theory level. 

Treating the Hubbard interaction futher in the RPA and 
Fourier transforming the susceptibility from imaginary time $\tau$  
to $(\vec k,i\omega)$ space, the susceptibility tensor then obeys
the Dyson equation,
\begin{eqnarray}
\tilde\chi^{RPA}  &=&  \tilde\chi^{(0)} +  U   \tilde\chi^{(0)}  \tilde\chi^{RPA} \, , 
\end{eqnarray}
which can be recast as,
\begin{eqnarray}
\tilde\chi^{RPA} &=& [\hat I-U \tilde\chi^{(0)}]^{-1} \tilde\chi^{(0)} \, .
\label{dyson}
\end{eqnarray}
Here $\hat I$ stands for the $2\times 2$ identity matrix.
Such a procedure of treating the interaction in RPA on top of the mean field
solution has been used in previous studies \cite{peres2002,brinckmann-lee}.  

\begin{figure}[hbt]
\begin{center}
\centerline{\includegraphics[width=6.00cm]{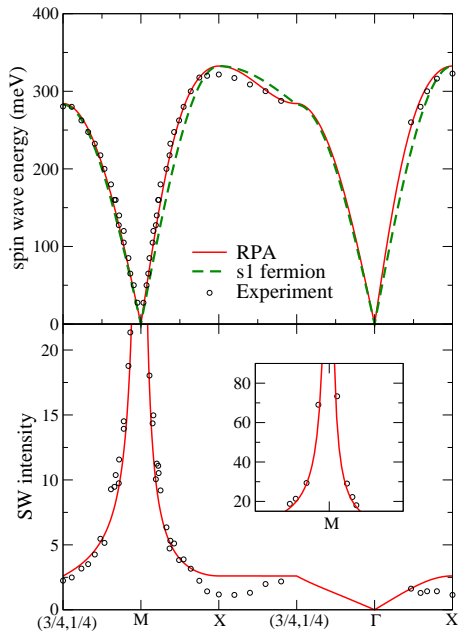}}
\caption{(Top) Spin-wave excitation spectrum along BZ special directions  
as specified in Ref. \cite{headings2010}.
(Bottom) Spin-wave intensity as obtained from the poles of the susceptybility (see text).
Experimental points from Ref. \cite{headings2010}.}
\label{fig2}
\end{center}
\end{figure}

$\tilde\chi^{RPA} $ has a pole $i\omega =\omega(\vec{k})$ obtained from the equation 
${\rm Det}\,[1-U \tilde\chi^{(0)}]=0$, which provides the dispersion relation $\omega(\vec{k})$ for the spin waves.
It has been shown in Ref. \cite{peres2002} that an excellent agreement with the spin-wave spectrum 
from Ref. \cite{LCO-2001} is achieved. In Fig. \ref{fig2} (top) we show a fit to the 
more recent experimental data of Ref. \cite{headings2010} (solid line) along with
the results from the $s1$ fermion method reported below in 
Sec. \ref{spin-spectrum-incoh} (dashed line) for $U/t=6.1$ and $t=295$\,meV. This
corresponds to a bandwidth $8t\approx 2.36$ eV. 

Importantly, provided that the $t$ magnitude is slightly increased for increasing values of $U/t$, 
agreement with the LCO spin-weight spectrum and distribution can be 
obtained for the range $U/t\in (6,8)$ and thus $U/8t\in (0.75,1)$ in units of the
bandwidth $8t$. For $U/t$ values smaller than $6$ (and larger than $8$), the spin-wave 
dispersion between $[\pi,0]$ and $[\pi/2,\pi/2]$ has a too large energy bandwidth (and is too flat) for any 
reasonable value of $t$. 

Let $\vert\nu,\omega (\vec{k})\rangle$ denote the excited energy eigenstates of energy 
$\omega (\vec{k})$ and momentum $\vec{k}$ that contribute to the coherent spin-wave spectral weight.
In the case of the $-+$ spin dynamical structure factor given in Eq. (\ref{SS}) for $\alpha =-$ and $\alpha' =+$,
the corresponding coherent spin-wave spectral weight in units of $\mu_B^2$ is
given by $Z_d\,2(1-2d)$. The factor $Z_d$ in this expression reads,
\begin{equation}
Z_d = 1 - {2\over N_a (1-2d)}\sum_{\vec{k}}\sum_{\nu'\neq \nu}\vert\langle\nu'\vert {\hat{s}}^{+}_{\vec{k},s}\vert GS\rangle\vert^2 \, ,
\label{Zd}
\end{equation}
where ${\hat{s}}^{+}_{\vec{k},s}$ is the Fourier transform of
the spin operator ${\hat{s}}^{+}_{\vec{r}_j,s}$ defined in Eq. (\ref{S-j-s})
and the sum over energy eigenstates excludes those that generate the coherent spin-wave 
weight, $\vert\nu\rangle = \vert\nu,\omega (\vec{k})\rangle$. 
In the $U/t\rightarrow\infty$ limit, $Z_d$ may be identified with the corresponding
$Z_d = Z_c\,Z_{\chi}$ factor of the Heisenberg model on the square lattice. 
According to the results of Ref. \cite{Canali}, the factors $Z_c$ and $Z_{\chi}$ have magnitudes
$Z_c\approx 1.18$ and  $Z_{\chi}\approx 0.48$, respectively, so that $Z_d \approx 0.57$. 
The limiting values $Z_d=1$ for $U/t\rightarrow 0$ and $Z_d \approx 0.57$ for $U/t\rightarrow\infty$
and the approximate intermediate value \cite{lorenzana2005}, $Z_d\approx 0.65$, at $U/t=8$ are recovered
as solutions of the equation,
\begin{equation} 
Z_d = e^{-Z_d\tanh \left(\sqrt{U\over 4\pi t}\right)} \, ,
\label{Z-d-app}
\end{equation}
which is used here for finite $U/t$. 

The spin dynamical structure factor measured in the high-energy inelastic neutron scattering
experiments of Ref. \cite{headings2010} includes a Bragg peak associated with its elastic part. The
corresponding elastic spectral weight is included in the total spin-weight sum-rule, $\mu_B^2\,2(1-2d)$,
of Eq. (\ref{sr-eff}). In the thermodynamic limit the upper-Hubbard band processes generate nearly no spin weight.
Hence the longitudinal spectral weight within the sum-rule $\mu_B^2\,2(1-2d)$ refers to the
elastic contribution. The elastic weight is given by $\approx \mu_B^2\,4(m_{AF})^2$. 
The inelastic spin spectral weight corresponds to the remaining weight in 
the spin-weight sum-rule $\approx \mu_B^2\,[2(1-2d)-4(m_{AF})^2]$. It refers to the $-+$ spin dynamical 
structure factor given in Eq. (\ref{SS}) for $\alpha =-$ and $\alpha' =+$.
It then follows that the experimentally determined spin-wave intensity, which corresponds to the coherent part of the inelastic spin
spectral weight, is in units of $\mu_B^2$ approximately given by,
\begin{equation}
W_{SW} \approx Z_d\,[2(1-2d)-4(m_{AF})^2] \, .
\label{W-SW}
\end{equation}

The GA+RPA method used in Ref. \cite{lorenzana2005} accounts for the quantum fluctuations that control 
the longitudinal and transverse relative weights. Within our description notations, that method is designed
to make the inelastic spin spectral weight  $\mu_B^2\,[2(1-2d)-4(m_{AF})^2]$ rather than 
$\mu_B^2\,2(1-2d)$. Hence for that GA+RPA method the spin-wave intensity factor is $Z_d$ as
defined in Eq. (\ref{Zd}). 

On the other hand, the RPA used here refers to the $-+$ spin dynamical structure factor alone. 
Hence it implicitly considers that the total inelastic spin spectral weight is $\mu_B^2\,2(1-2d)$ rather than 
$\mu_B^2\,[2(1-2d)-4(m_{AF})^2]$. Therefore, to describe the actual spin-wave intensity momentum 
distribution one must use a corresponding experimentally determined factor $Z_d^{exp}<Z_d$ such that, 
\begin{equation}
W_{SW} = Z_d^{exp}\,2(1-2d) = Z_d\,[2(1-2d)-4(m_{AF})^2] \, .
\label{W-SW-exp}
\end{equation}

From the residue of the spin-wave pole the susceptibility coherent part then reads,
\begin{equation}
\chi_{co}^{-+} (\vec{k},i\omega) = Z_d^{exp}\sum_{l=\pm1}
\frac{{\rm Res}\,[\chi^{-+}(\vec{k},l\,\omega(\vec{k})] }{i\omega-l\,\omega(\vec{k})} \, ,
\label{completechi}
\end{equation}
with $\chi^{-+}$ obtained in RPA above. The measured intensity is \cite{authors}, 
\begin{eqnarray}
I_{SW}(\vec{k})=\pi [S^{xx}(\vec{k})+S^{yy}(\vec{k})]=\pi S^{-+}(\vec{k})\,.
\end{eqnarray}

In Fig. \ref{fig2} (bottom), we plot the corresponding RPA spin-wave intensity, 
\begin{equation}
I_{SW}(\vec{k})=-[\pi/2] Z_d^{exp}{\rm Res}\,[\chi^{-+} (\vec{k},\omega(\vec{k})] \, .
\end{equation}
The good agreement with the experimental data, specially near the point $M$, reproduces the theoretical 
results of Ref. \cite{headings2010}. It is here obtained for the value $Z_d^{exp}\approx 0.49$,
which corresponds to the choice $m_{AF}=m_{AF}^{GB}= m_{AF}^{0}(1-2d)$ such that
$\delta S\approx d$. The $m_{AF}^{GB}$ dependence on $U/t$ is given in Eq. (\ref{M-AF-limits}) for $U/t<8$.
As in Ref. \cite{headings2010}, the spin-wave intensity shows disagreement around the $X$ point, which here 
probably stems from effects not captured by the RPA.

\section{The rotated-electron description emerging from the model's extended global symmetry}
\label{general-rep}

The goal of this section is the introduction of the general rotated-electron
representation from which the spinon representation used in the ensuing section 
naturally emerges. In contrast to the remaining sections of this paper, here we consider 
arbitrary values of the electronic density $n=N/N_a$ and spin density 
$m=[N_{\uparrow}-N_{\downarrow}]$. 

\subsection{The electron - rotated-electron unitary operator}
\label{rot-elec}

We denote the spin and $\eta$-spin of an energy eigenstate by $S_s$ and
$S_{\eta}$, respectively. The corresponding spin and $\eta$-spin projections 
read $S_{s}^z=-{1\over 2}[N_{\uparrow}-N_{\downarrow}]$ and
$S_{\eta}^z=-{1\over 2}[N_a-N]$, respectively. The lowest-weight states (LWSs) of 
both the $\eta$-spin and spin algebras are such that $S_{\alpha}= -S_{\alpha}^{z}$ 
where  $\alpha =\eta$ for $\eta$-spin and $\alpha =s$ for spin. The numbers,
\begin{eqnarray}
n_{\eta} & = & S_{\eta}-{1\over 2}(N_a-N) = 0,1,..., 2S_{\eta} \, ,
\nonumber \\
n_{s} & = & S_{s}-{1\over 2}(N_{\uparrow}-N_{\downarrow}) = 0,1,..., 2S_{s} \, ,
\label{n-eta.n-s}
\end{eqnarray}
vanish for such a LWS.

Let $\{\vert \Psi_{l_r,l_{\eta s},u}\rangle\}$ be a complete set of $4^{N_a}$ energy, 
momentum, $\eta$-spin, $\eta$-spin projection, spin, and spin-projection eigenstates for
$u\equiv U/t>0$. Here $l_{\eta s}$ is a short notation for the set of four quantum numbers $[S_{\eta},S_{s},n_{\eta},n_s]$ 
and the index $l_r$ represents all remaining quantum numbers, other than 
those, that are needed to fully specify an energy eigenstate $\vert \Psi_{l_r,l_{\eta s},u}\rangle$. 
The energy eigenstates of that set that are not LWSs are generated from those as follows,
\begin{equation}
\vert \Psi_{l_r,l_{\eta s},u}\rangle = \prod_{\alpha=\eta,s}\left[\frac{1}{
\sqrt{{\cal{C}}_{\alpha}}}({\hat{S}}^{\dag}_{\alpha})^{n_{\alpha}}\right]\vert \Psi_{l_r,l_{\eta s}^0,u}\rangle \, .
\label{Gstate-BAstate}
\end{equation}
Here, 
\begin{eqnarray}
{\cal{C}}_{\alpha} & = & \langle \Psi_{l,l_{\eta s}^0,u}\vert ({\hat{S}}_{\alpha})^{n_{\alpha}}({\hat{S}}^{\dag}_{\alpha})^{n_{\alpha}}\vert \Psi_{l,l_{\eta s}^0,u}\rangle
\nonumber \\
& = & [n_{\alpha}!]\prod_{j'=1}^{n_{\alpha}}[\,2S_{\alpha}+1-j'\,] \, , 
\hspace{0.25cm} \alpha = \eta,s  \, ,
\label{Calpha}
\end{eqnarray}
for $n_{\alpha}=1,...,2S_{\alpha}$ are normalization constants, the $\eta$-spin ($\alpha =\eta$) and spin ($\alpha =s$) off-diagonal generators 
${\hat{S}}^{\dag}_{\alpha}$ and ${\hat{S}}_{\alpha}$ are given in Eq. (\ref{Scs}) of Appendix \ref{commute}, and $l_{\eta s}$ and 
$l_{\eta s}^0$ stand for $[S_{\eta},S_{s},n_{\eta},n_s]$
and $[S_{\eta},S_{s},0,0]$, respectively. Within our notation, $l_{\eta s}^0$
refers to values of the general index $l_{\eta s}$ associated with a LWS
such that $n_{\eta}= n_s=0$. 

For the Hubbard model on the square lattice and also on the 1D lattice,
upon adiabatically increasing $U/t$ from any finite value to the $U/t\rightarrow\infty$ limit, 
each energy eigenstate $\vert \Psi_{l_r,l_{\eta s},u}\rangle$ continuously
evolves into a uniquely defined corresponding energy eigenstate $\vert \Psi_{l_r,l_{\eta s},\infty}\rangle$,
and vice versa. We emphasize though that due to the high degeneracy among different spin sectors as
well as $\eta$-spin sectors that occurs in the $U/t\rightarrow\infty$ limit, there are in such
a limit many more choices of energy eigenstates sets than for $U/t$ finite. Accordingly,
upon adiabatically decreasing $U/t$ most of such $U/t\rightarrow\infty$ states
do not evolve into finite-$U/t$ energy eigenstates. Our above
procedure uniquely defines a convenient set of $U/t\rightarrow\infty$ 
energy eigenstates that upon adiabatically decreasing $U/t$ do 
evolve into finite-$U/t$ energy eigenstates.

Both the corresponding sets of $4^{N_a}$ states $\{\vert \Psi_{l_r,l_{\eta s},u}\rangle\}$ and  $\{\vert \Psi_{l_r,l_{\eta s},\infty}\rangle\}$, respectively, 
are complete and refer to the same Hilbert space. Hence there is a uniquely defined unitary transformation connecting the states
$\vert \Psi_{l_r,l_{\eta s},u}\rangle$ and $\vert \Psi_{l_r,l_{\eta s},\infty}\rangle$.
Indeed, since the model's Hilbert space is the same for all $U/t>0$ values considered here,
it follows from basic quantum-mechanics Hilbert-space and operator properties that
for this choice there exists exactly one unitary operator ${\hat{V}}={\hat{V}}(U/t)$
such that any $U/t\rightarrow\infty$ energy eigenstate $\vert \Psi_{l_r,l_{\eta s},\infty}\rangle$
is transformed onto the corresponding $U/t>0$ energy eigenstate $\vert \Psi_{l_r,l_{\eta s},u}\rangle$ as,
\begin{equation}
\vert \Psi_{l_r,l_{\eta s},u}\rangle ={\hat{V}}^{\dag}\vert \Psi_{l_r,l_{\eta s},\infty}\rangle \, .
\label{Psi-V-Psi}
\end{equation}
The energy eigenstates $\vert \Psi_{l_r,l_{\eta s},u}\rangle ={\hat{V}}^{\dag}\vert \Psi_{l_r,l_{\eta s},\infty}\rangle$
(one for each value of $U/t>0$) that are generated from the same initial 
$U/t\rightarrow\infty$ energy eingenstate $\vert \Psi_{l_r,l_{\eta s},\infty}\rangle$ 
belong to the same {\it $V$ tower}. 

The rotated-electron operators are given by,
\begin{eqnarray}
{\tilde{c}}_{\vec{r}_j,\sigma}^{\dag} & = &
{\hat{V}}^{\dag}\,c_{\vec{r}_j,\sigma}^{\dag}\,{\hat{V}}
\, ; \hspace{0.35cm}
{\tilde{c}}_{\vec{r}_j,\sigma} =
{\hat{V}}^{\dag}\,c_{\vec{r}_j,\sigma}\,{\hat{V}} \, ,
\nonumber \\
{\tilde{n}}_{\vec{r}_j,\sigma} & = & {\tilde{c}}_{\vec{r}_j,\sigma}^{\dag}\,{\tilde{c}}_{\vec{r}_j,\sigma} 
\, ; \hspace{0.35cm} {\hat{V}} = e^{-{\hat{S}}} \, .
\label{rotated-operators}
\end{eqnarray}
For $U/t>0$ the operator $\hat{S}$ appearing here can be expanded in a series of $t/U$
whose leading-order term is provided in Eq. (\ref{OOr}) of Appendix \ref{commute}.

Since the electron - rotated-electron unitary operator ${\hat{V}}$ commutes with itself, the
equalities ${\hat{V}} = e^{-{\hat{S}}}={\tilde{V}} = e^{-{\tilde{S}}}$ and
${\hat{S}}={\tilde{S}}$ hold. Hence both the operators
${\hat{V}}$ and ${\hat{S}}$ have the same expression in terms of
electron and rotated-electron creation and annihilation operators. It then
follows from the expression of the operator ${\hat{S}}$ provided in Eq.
(\ref{OOr}) of Appendix \ref{commute} that the corresponding rotated
operator ${\tilde{S}}$ has the following leading-order term,
\begin{equation}
{\tilde{S}} = -{t\over U}\,\left[\tilde{T}_{+1} -\tilde{T}_{-1}\right] + ... \, .
\label{Sr-rot}
\end{equation} 
The rotated kinetic operators $\tilde{T}_{+1}$ and $\tilde{T}_{-1}$ appearing here
and the related rotated kinetic operator $\tilde{T}_0$ are given in Eq. (\ref{T-op-rot})
of Appendix \ref{commute}. The expressions of the corresponding unrotated kinetic operators 
$\hat{T}_0$, $\hat{T}_{+1}$, and $\hat{T}_{-1}$ are provided in Eq. (\ref{T-op}) of that Appendix.

Note that the equality ${\hat{S}}={\tilde{S}}$ refers to the whole expression of these operators. 
An important property for our study is that except in the
$U/t\rightarrow\infty$ limit the leading order terms $-{t\over U}\,[\hat{T}_{+1} -\hat{T}_{-1}]$ 
and $-{t\over U}\,[\tilde{T}_{+1} -\tilde{T}_{-1}]$ of the operators ${\hat{S}}$ and ${\tilde{S}}$
given in Eq. (\ref{OOr}) of Appendix \ref{commute} and Eq. (\ref{Sr-rot}), respectively,
are different operators. Moreover, except for $U/t\rightarrow\infty$ one has that $\hat{T}_0\neq \tilde{T}_0$, 
$\hat{T}_{+1}\neq \tilde{T}_{+1}$, and $\hat{T}_{-1}\neq \tilde{T}_{-1}$.
This is behind for intermediate $U/t$ values the few first terms of the Hamiltonian $t/U$
expansion as written in terms of rotated-electron operators containing much more
complicated higher-order terms when expressed in terms of electron creation and
annihilation operators.

The main point here is that for the rotated electrons that emerge from the unitary
transformation of Eq. (\ref{rotated-operators}) single and double occupancy are good quantum
numbers for $U/t>0$. Indeed, on any bipartite lattice the number of rotated-electron singly occupied sites operator,
\begin{eqnarray}
2{\tilde{S}}_{c} & = & {\hat{V}}^{\dag}\,{\hat{Q}}\,{\hat{V}} =\sum_{j=1}^{N_a}{\tilde{s}}_{\vec{r}_j,c} \, ,
\nonumber \\
{\tilde{s}}_{\vec{r}_j,c} & = & {\hat{V}}^{\dag}\,{\hat{s}}_{\vec{r}_j,c}\,{\hat{V}} = \sum_{\sigma =\uparrow
,\downarrow}\,{\tilde{n}}_{\vec{r}_j,\sigma}\,(1- {\tilde{n}}_{\vec{r}_j,-\sigma}) \, ,
\label{Or-ope}
\end{eqnarray}
commutes with the Hubbard model Hamiltonian \cite{bipartite}. Here ${\hat{Q}}$ is the corresponding 
number of electron singly occupied sites operator given in Eq. (\ref{DQ}).
This follows in part from the symmetries of the Hamiltonian electron-interaction term,
which imply that all $U/t\rightarrow\infty$ energy eigenstates of the set $\{\vert \Psi_{l_r,l_{\eta s},\infty}\rangle\}$ are
as well eigenstates of the electron double-occupancy operator ${\hat{D}}$ and single-occupancy operator
${\hat{Q}}$ provided in that equation. Hence in the $U/t\rightarrow\infty$ limit the Hilbert space
is classified in subspaces with different numbers of doubly-occupied sites
and each of the states $\{\vert \Psi_{l_r,l_{\eta s},\infty}\rangle\}$ is contained in only one
of these subspaces. The same applies to the $4^{N_a}$ energy eigenstates of the set
$\{\vert \Psi_{l_r,l_{\eta s},u}\rangle\}$ for $U/t>0$ in terms of rotated-electron doubly-occupied sites.

The unitary operator ${\hat{V}}$ of our formulation is uniquely defined by its $4^{N_a}\times 4^{N_a}$ matrix
elements, $\langle\Psi_{l_r,l_{\eta s},u}\vert {\hat{V}}\vert \Psi_{l_r',l_{\eta s}',u}\rangle$.
For $U/t>0$ most of these matrix elements vanish. For $U/t\rightarrow\infty$
rotated electrons become electrons so that the matrix representing the unitary operator ${\hat{V}}$
becomes the $4^{N_a}\times 4^{N_a}$ unit matrix. Hence
$\langle\Psi_{l_r,l_{\eta s},\infty}\vert {\hat{V}}\vert \Psi_{l_r',l_{\eta s}',\infty}\rangle=\delta_{l_r,l_r'}\delta_{l_{\eta s},l_{\eta s}'}$. 
On the other hand, as justified in Appendix \ref{commute},
the unitary operator ${\hat{V}}={\hat{V}}(U/t)$ commutes with the six generators of the global 
$\eta$-spin and spin $SU(2)$ symmetries. This implies that the matrix elements between energy eigenstates 
with different values of $S_{\eta}$, $S_s$, $n_{\eta}$, and $n_s$ and thus of $l_{\eta s}$ vanish.
Hence the finite matrix elements are between states with the same $l_{\eta s}$ values
so that we denote them by $V_{l_r,l_r'}$,
\begin{equation}
\langle\Psi_{l_r,l_{\eta s},\infty}\vert {\hat{V}}\vert \Psi_{l_r',l_{\eta s}',\infty}\rangle
= \delta_{l_{\eta s},l_{\eta s}'}V_{l_r,l_r'} \, ,
\label{ME-GEN}
\end{equation}
where,
\begin{eqnarray}
V_{l_r,l_r'} & = & \langle\Psi_{l_r,l_{\eta s},u}\vert {\hat{V}}\vert \Psi_{l_r',l_{\eta s},u}\rangle
= \langle\Psi_{l_r',l_{\eta s},\infty}\vert {\hat{V}}^{\dag}\vert \Psi_{l_r,l_{\eta s},\infty}\rangle^*
\nonumber \\
& = & \langle\Psi_{l_r,l_{\eta s},u}\vert \Psi_{l_r',l_{\eta s},\infty}\rangle = 
\langle\Psi_{l_r',l_{\eta s},\infty}\vert \Psi_{l_r,l_{\eta s},u}\rangle^* \, .
\label{ME-Vll}
\end{eqnarray}

Given a complete set of $4^{N_a}$ energy, momentum, $\eta$-spin, $\eta$-spin projection, spin, 
and spin-projection eigenstates, $\{\vert \Psi_{l_r,l_{\eta s},u}\rangle\}$, the electron - rotated-electron
unitary operator considered here is for $U/t>0$ uniquely defined by the matrix elements of
Eqs. (\ref{ME-GEN}) and (\ref{ME-Vll}). This corresponds to one out of the infinite choices
of electron - rotated-electron unitary operators \cite{bipartite}. 
All these operators and corresponding unitary transformations
refer to the same subspaces with fixed numbers of doubly-occupied sites, $0,1,2,3,...$.
They only differ in the choice of basis states within each of these subspaces. 
For most of these unitary operators the states ${\hat{V}}^{\dag}\vert \Psi_{l_r,l_{\eta s},\infty}\rangle$
are not energy eigenstates for finite $U/t$ values. The electron - rotated-electron unitary 
transformation considered here has been constructed to  make  these states
energy eigenstates for finite $U/t$ values, as given in Eq. (\ref{Psi-V-Psi}).

\subsection{The general operational description naturally emerging from 
the rotated electrons and symmetry}
\label{description}

The electron - rotated-electron unitary transformation is closely related to the extended global $SO(3)\times SO(3)\times U(1)$
symmetry found in Ref. \cite{bipartite} for the Hamiltonian given in Eq. (\ref{HH}) on any bipartite lattice. 
Until recently \cite{Zhang} it was believed that the model's global  symmetry was for finite on-site interaction values only 
$SO(4) =[SU(2)\otimes SU(2)]/Z_2$. The occurrence of a global $c$ hidden $U(1)$ symmetry beyond $SO(4)$
in the model's global $SO(3)\otimes SO(3)\otimes U(1)=[SO(4)\otimes U(1)]/Z_2$ symmetry must be accounted
for in studies of the Hubbard model on any bipartite lattice. Such a global symmetry 
may be rewritten as $[SU(2)\times SU(2)\times U(1)]/Z_2^2$ and stems from the $U\neq 0$ 
local gauge $SU(2)\times SU(2) \times U(1)$ symmetry of the Hubbard model on a bipartite lattice with vanishing 
transfer integral, $t=0$ \cite{U(1)-NL}. The seven local generators of the corresponding two gauge $SU(2)$ symmetries and $U(1)$ symmetry
are the three spin local operators ${\hat{s}}_{\vec{r}_j,s}^{l}$ provided in Eq. (\ref{S-j-s}) and the three $\eta$-spin local operators
${\hat{s}}_{\vec{r}_j,\eta}^{l}$ and the local operator ${\hat{s}}_{\vec{r}_j,c}$ given in Eqs. (\ref{S-j-eta}) and (\ref{S-j-c}) of Appendix \ref{commute},
respectively. The index $l$ in the generators of the two $SU(2)$ symmetries stand for $l =\pm, z$.

An important point is that although addition of chemical-potential and magnetic-field operator terms to the Hubbard 
model on a square lattice Hamiltonian given in Eq. (\ref{HH}) lowers its symmetry, these terms commute with it. 
Therefore, the global symmetry of the latter Hamiltonian being $SO(3)\otimes SO(3)\otimes U(1)$ implies that the set 
of independent rotated-electron occupancy configurations that generate all $4^{N_a}$ energy eigenstates,
$\{\vert \Psi_{l_r,l_{\eta s},u}\rangle\}$, generate as well representations of the global symmetry algebra for all 
values of electronic density $n$ and spin density $m$. It is confirmed in Ref. \cite{bipartite} that the
number of these independent $[SU(2)\otimes SU(2)\otimes U(1)]/Z_2^2=SO(3)\otimes SO(3)\otimes U(1)$
symmetry algebra representations equals for the present model on a bipartite lattice its Hilbert-space dimension, $4^{N_a}$.

The generator ${\hat{s}}_{\vec{r}_j,c}$ of the local gauge $U(1)$ symmetry given in Eq. (\ref{S-j-c}) of Appendix \ref{commute}
and the alternative local generator ${\hat{s}}_{\vec{r}_j,c}^h = (1 - {\hat{s}}_{\vec{r}_j,c})$ may be expressed as,
\begin{eqnarray}
{\hat{s}}_{\vec{r}_j,c} & = & {\hat{q}}_{\vec{r}_j}^c \equiv {\hat{f}}_{\vec{r}_j,c}^{\dag}\,{\hat{f}}_{\vec{r}_j,c} \, ,
\nonumber \\
{\hat{s}}_{\vec{r}_j,c}^h & = & (1-{\hat{q}}_{\vec{r}_j}^c) = {\hat{f}}_{\vec{r}_j,c}\,{\hat{f}}_{\vec{r}_j,c}^{\dag} \, .
\label{n-r-c}
\end{eqnarray}
Here ${\hat{f}}_{\vec{r}_j,c}^{\dag}$ and ${\hat{f}}_{\vec{r}_j,c}$ stand for the following creation and annihilation
operators, respectively, of suitable spin-less and $\eta$-spin-less fermions, 
\begin{eqnarray}
{\hat{f}}_{\vec{r}_j,c}^{\dag} & = & c_{\vec{r}_j,\uparrow}^{\dag}\,(1-{\hat{n}}_{\vec{r}_j,\downarrow})
+ e^{i\vec{\pi}\cdot\vec{r}_j}\,c_{\vec{r}_j,\uparrow}\,{\hat{n}}_{\vec{r}_j,\downarrow} \, ,
\nonumber \\
{\hat{f}}_{\vec{r}_j,c} & = & c_{\vec{r}_j,\uparrow}\,(1-{\hat{n}}_{\vec{r}_j,\downarrow})
+ e^{i\vec{\pi}\cdot\vec{r}_j}\,c_{\vec{r}_j,\uparrow}^{\dag}\,{\hat{n}}_{\vec{r}_j,\downarrow}  \, ,
\label{fc-unrot}
\end{eqnarray}
where we used that $e^{i\vec{\pi}\cdot\vec{r}_j}=e^{-i\vec{\pi}\cdot\vec{r}_j}$. Here and
throughout this paper the vector $\vec{\pi}$ has Cartesian components $\vec{\pi}=[\pi,\pi]$. 

We call {\it $c$ fermions} the rotated-electron
related objects whose creation and annihilation operators 
$f_{\vec{r}_j,c}^{\dag} ={\hat{V}}^{\dag}\,{\hat{f}}_{\vec{r}_j,c}^{\dag}\,{\hat{V}}$ and $f_{\vec{r}_j,c} ={\hat{V}}^{\dag}\,{\hat{f}}_{\vec{r}_j,c}\,{\hat{V}}$, 
respectively, are generated from those of the spin-less and $\eta$-spin-less fermions of Eq. (\ref{fc-unrot})
by the specific electron - rotated-electron unitary transformation uniquely defined by the matrix elements of
Eqs. (\ref{ME-GEN}) and (\ref{ME-Vll}). (No 
upper index $\tilde{f}$ is used onto the (rotated) $c$ fermion operator $f_{\vec{r}_j,c}$.) Hence these operators read,
\begin{eqnarray}
f_{\vec{r}_j,c}^{\dag} & = & {\tilde{c}}_{\vec{r}_j,\uparrow}^{\dag}\,
(1-{\tilde{n}}_{\vec{r}_j,\downarrow}) + e^{i\vec{\pi}\cdot\vec{r}_j}\,{\tilde{c}}_{\vec{r}_j,\uparrow}\,{\tilde{n}}_{\vec{r}_j,\downarrow} \, ,
\nonumber \\
f_{\vec{r}_j,c} & = & {\tilde{c}}_{\vec{r}_j,\uparrow}\,
(1-{\tilde{n}}_{\vec{r}_j,\downarrow}) + e^{i\vec{\pi}\cdot\vec{r}_j}\,{\tilde{c}}_{\vec{r}_j,\uparrow}^{\dag}\,{\tilde{n}}_{\vec{r}_j,\downarrow} \, .
\label{fc+}
\end{eqnarray}
The rotated-electron creation and annihilation operators appearing here are generated from corresponding electron
operators by the unitary transformation uniquely defined above, as given in Eq.
(\ref{rotated-operators}). The corresponding $c$ fermion local density operator is given by,
\begin{equation}
{\tilde{q}}_{\vec{r}_j}^c= f_{\vec{r}_j,c}^{\dag}\,f_{\vec{r}_j,c} \, .
\label{n-r-c-rot}
\end{equation}

The $c$ fermions live on a lattice identical to the original lattice.
One can introduce $c$ fermion momentum dependent operators \cite{companion,companion0},
\begin{equation}
f_{\vec{q}_j,c}^{\dag} =
{1\over {\sqrt{N_a}}}\sum_{j'=1}^{N_a}\,e^{+i\vec{q}_j\cdot\vec{r}_{j'}}\,
f_{\vec{r}_{j'},c}^{\dag} \, ; \hspace{0.35cm} j=1,...,N_a \, .
\label{fc+q}
\end{equation}
Here the $c$ fermion operators $f_{\vec{r}_{j'},c}^{\dag}$ where the index $j'=1,...,N_a$
refers to the sites of the original lattice are mapped from the rotated-electron 
operators by an exact local transformation given in
Eq. (\ref{fc+}). The $c$ momentum band has $N_a$ discrete momentum values 
$\vec{q}_j$ where $j=1,...,N_a$. 
It has the same shape and momentum area as the electronic first-BZ.

The generator $2{\tilde{S}}_c$ of the related global $c$ hidden $U(1)$ symmetry in $[SU(2)\times SU(2)\times U(1)]/Z_2^2$ 
found in Ref. \cite{bipartite} is the number of rotated-electron singly occupied sites given in Eq. (\ref{Or-ope}).
Hence it involves the site summation $\sum_{j=1}^{N_a}$ over the rotated local generator
${\tilde{s}}_{\vec{r}_j,c}$ rather than over the corresponding unrotated local operator
${\hat{s}}_{\vec{r}_j,c}$ of Eq. (\ref{S-j-c}) of Appendix \ref{commute}. 
This is why $2{\tilde{S}}_c= {\hat{V}}^{\dag}\,{\hat{Q}}\,{\hat{V}}$,
as given in Eq. (\ref{Or-ope}), where the operator ${\hat{Q}}$ is that of Eq. (\ref{DQ}). The eigenvalues 
$2S_c=0,1,2,...$ of the generator $2{\tilde{S}}_c$ are thus the numbers of rotated-electron singly occupied sites. 

The $c$ fermion creation and annihilation operators are found in Appendix
\ref{commute} to obey the anti-commutation relations given in Eq. (\ref{albegra-cf}) of that Appendix. A straightforward 
operator algebra then confirms that the $c$ fermion local 
density operator of Eq. (\ref{n-r-c-rot}) is the local operator ${\tilde{s}}_{\vec{r}_j,c}$ appearing in the
expression provided in Eq. (\ref{Or-ope}). Hence the global $c$ hidden $U(1)$ symmetry generator may be 
simply rewritten as,
\begin{equation}
2{\tilde{S}}_c = \sum_{j=1}^{N_a}{\tilde{q}}_{\vec{r}_j}^c= \sum_{j=1}^{N_a} f_{\vec{r}_j,c}^{\dag}\,f_{\vec{r}_j,c} \, .
\label{Or-ope-cf}
\end{equation}

One finds that except in the $U/t\rightarrow\infty$ limit the inequality $\sum_{j=1}^{N_a}{\hat{q}}_{\vec{r}_j}^c \neq  
\sum_{j=1}^{N_a}{\tilde{q}}_{\vec{r}_j}^c$ holds. This confirms that the generator $2{\tilde{S}}_c$ given in
Eq. (\ref{Or-ope-cf}) does not commute with the electron - rotated-electron unitary operator ${\hat{V}}={\tilde{V}}$.
On the other hand and as justified in Appendix \ref{commute}, the three components of the momentum 
operator $\hat{{\vec{P}}}$, three generators of the global spin $SU(2)$ symmetry, and three generators
of the global $\eta$-spin $SU(2)$ symmetry commute with that 
unitary operator. Hence in contrast to the Hamiltonian and generator $2{\tilde{S}}_c$, these operators have
the same expression in terms of electron and rotated-electron 
creation and annihilation operators, as given in Eqs.
(\ref{P-invariant}) and (\ref{Scs}) of Appendix \ref{commute}. 
On the contrary, the generator of the global $c$ hidden $U(1)$ symmetry
given in Eq. (\ref{Or-ope}) does not commute with the unitary operator
${\hat{V}}$. This is behind the hidden character of such a symmetry.

Site summation $\sum_{j=1}^{N_a}$ over the rotated local operator provided in 
Eq. (\ref{n-r-c-rot}) and over the following six rotated local operators,
\begin{eqnarray}
{\tilde{s}}_{\vec{r}_j,\eta}^{z} & = & -{1\over 2}[1-{\tilde{n}}_{\vec{r}_j,\uparrow}-{\tilde{n}}_{\vec{r}_j,\downarrow}] \, ,
\nonumber \\
{\tilde{s}}_{\vec{r}_j,\eta}^{+} & = & e^{i\vec{\pi}\cdot\vec{r}_j}\,{\tilde{c}}_{\vec{r}_j,\downarrow}^{\dag}\,
{\tilde{c}}_{\vec{r}_j,\uparrow}^{\dag} \, ,
\nonumber \\
{\tilde{s}}_{\vec{r}_j,\eta}^{-} & = & e^{-i\vec{\pi}\cdot\vec{r}_j}\,{\tilde{c}}_{\vec{r}_j,\uparrow}\,{\tilde{c}}_{\vec{r}_j,\downarrow} \, ,
\nonumber \\
{\tilde{s}}_{\vec{r}_j,s}^{z} & = & -{1\over 2}[{\tilde{n}}_{\vec{r}_j,\uparrow}-{\tilde{n}}_{\vec{r}_j,\downarrow}] \, ,
\nonumber \\
{\tilde{s}}_{\vec{r}_j,s}^{+} & = & {\tilde{c}}_{\vec{r}_j,\downarrow}^{\dag}\,{\tilde{c}}_{\vec{r}_j,\uparrow} \, ,
\nonumber \\
{\tilde{s}}_{\vec{r}_j,s}^{-} & = & {\tilde{c}}_{\vec{r}_j,\uparrow}^{\dag}\,
{\tilde{c}}_{\vec{r}_j,\downarrow} \, , \hspace{0.35cm}
j = 1,2,...,N_a  \, ,
\label{Scs-j-rot}
\end{eqnarray}
gives the seven generators of the model's global 
$SO(3)\otimes SO(3)\otimes U(1)=[SU(2)\otimes SU(2)\otimes U(1)]/Z_2^2$ symmetry,
as provided in Eqs. (\ref{Or-ope}) and Eq. (\ref{Scs}) of Appendix \ref{commute}. 
However, except in the $U/t\rightarrow\infty$ 
limit the six rotated local operators given in Eq. (\ref{Scs-j-rot}) and the corresponding six unrotated local operators 
provided in Eq. (\ref{S-j-s}) and Eq. (\ref{S-j-eta}) of Appendix \ref{commute} are different operators.  

Interestingly, the $\eta$-spin and spin $SU(2)$ symmetries are within the present representation particular 
cases of a general $\eta s$ quasi-spin $SU(2)$ symmetry. The corresponding three local 
$\eta s$ quasi-spin operators ${\tilde{q}}^l_{\vec{r}_j}$ such that $l =\pm, z$ obey a $SU(2)$ algebra
and have the following expression in terms of rotated-electron operators,
\begin{eqnarray}
{\tilde{q}}^-_{\vec{r}_j} & = & 
({\tilde{c}}_{\vec{r}_j,\uparrow}^{\dag}
+ e^{i\vec{\pi}\cdot\vec{r}_j}\,{\tilde{c}}_{\vec{r}_j,\uparrow})\,
{\tilde{c}}_{\vec{r}_j,\downarrow} \, ,
\nonumber \\
{\tilde{q}}^+_{\vec{r}_j} & = & ({\tilde{q}}^-_{\vec{r}_j})^{\dag} \, ;
\hspace{0.35cm}
{\tilde{q}}^{z}_{\vec{r}_j} = ({\tilde{n}}_{\vec{r}_j,\downarrow} - 1/2) \, .
\label{rotated-quasi-spin}
\end{eqnarray}
Here ${\tilde{q}}^{\pm}_{\vec{r}_j}= {\tilde{q}}^{x}_{\vec{r}_j}\pm i\,{\tilde{q}}^{y}_{\vec{r}_j}$ where $x,y,z$ denotes the Cartesian coordinates. 
The relation of these $\eta s$ quasi-spin operators to the original electron creation and annihilation operators involves
the unitary transformation of Eq. (\ref{rotated-operators}).

Within the present rotated-electron operational formulation, three related elementary objects naturally emerge
that make the model's global symmetry explicit. The operators provided in Eq. (\ref{fc+}) create and annihilate 
spin-less and $\eta$-spin-less $c$ fermions whose local density operator, Eq. (\ref{n-r-c-rot}), is directly related to the
generator of the global $c$ hidden $U(1)$ symmetry, as given in Eq. (\ref{Or-ope-cf}). The $c$ fermions
carry the charges of the rotated electrons that singly occupy sites. Moreover,
the three rotated local spin operators ${\tilde{s}}^l_{\vec{r}_j,s}$ and the three rotated local $\eta$-spin  
operators ${\tilde{s}}^l_{\vec{r}_j,\eta}$ such that $l=\pm,z$ given in Eq. (\ref{Scs-j-rot}) are associated with the spin-$1/2$ spinons
and $\eta$-spin-$1/2$ $\eta$-spinons, respectively, as defined here. The spin-$1/2$ spinons carry the spin
of the rotated electrons that singly occupy sites. The $c$ fermion holes describe the degrees of freedom
associated with the $c$ hidden $U(1)$ symmetry of the sites doubly occupied and unoccupied by
the rotated electrons. The $\eta$-spin degrees of freedom of these sites are described by the $\eta$-spin
projection $-1/2$ $\eta$-spinons (rotated-electron doubly occupied sites) and $\eta$-spin
projection $+1/2$ $\eta$-spinons (rotated-electron unoccupied sites). 

Within our representation, the local operators ${\tilde{s}}_{\vec{r}_j,c}$, ${\tilde{s}}_{\vec{r}_j,c}^h$, and ${\tilde{s}}^l_{\vec{r}_j,\alpha}$ where $l=\pm,z$ 
and $\alpha =s,\eta$ can be expressed in terms of only the $c$ fermion local density operator ${\tilde{q}}_{\vec{r}_j}^c$ given in Eq. (\ref{n-r-c-rot})
and three local $\eta s$ quasi-spin operators ${\tilde{q}}^{l}_{\vec{r}_j}$ of Eq. (\ref{rotated-quasi-spin}) as follows,
\begin{eqnarray}
{\tilde{s}}_{\vec{r}_j,c} & = & {\tilde{q}}_{\vec{r}_j}^c\, ; \hspace{0.50cm} {\tilde{s}}_{\vec{r}_j,c}^h =  (1-{\tilde{q}}_{\vec{r}_j}^c) \, ,
\nonumber \\
{\tilde{s}}^l_{\vec{r}_j,s} & = & {\tilde{q}}_{\vec{r}_j}^c\,{\tilde{q}}^l_{\vec{r}_j} \, ; \hspace{0.50cm}
{\tilde{s}}^l_{\vec{r}_j,\eta} = (1-{\tilde{q}}_{\vec{r}_j}^c)\,{\tilde{q}}^l_{\vec{r}_j}  \, ,
\nonumber \\
l & = & \pm,z \, .
\label{sir-pir}
\end{eqnarray}
The expressions of the local spinon operators ${\tilde{s}}^l_{\vec{r}_j,s}$ and 
local $\eta$-spinon operators ${\tilde{s}}^l_{\vec{r}_j,\eta}$ provided here are a confirmation that the
corresponding spin $SU(2)$ and $\eta$-spin $SU(2)$ symmetries
are particular cases of the $\eta s$ quasi-spin $SU(2)$ symmetry. Specifically, they are associated with the
$SU(2)$ algebra representations involving the (i) spin-up and spin-down
rotated-electron singly occupied sites and (ii) rotated-electron doubly-occupied
and unoccupied sites, respectively. Indeed, the $c$ fermion and $c$ fermion
hole local density operators ${\tilde{q}}_{\vec{r}_j}^c$ and $(1-{\tilde{q}}_{\vec{r}_j}^c)$ play in
the expressions of these operators provided in Eq. (\ref{sir-pir}) the role
of projectors onto such two sets of lattice-site rotated-electron occupancies, respectively.

The relations given in Eq. (\ref{sir-pir}) for the operators ${\tilde{s}}^l_{\vec{r}_j,s}$ and ${\tilde{s}}^l_{\vec{r}_j,\eta}$ 
are equivalent to the following expression of the local $\eta s$ quasi-spin operators ${\tilde{q}}^{l}_{\vec{r}_j}$ 
in terms of those of the former operators provided in Eq. (\ref{Scs-j-rot}),
\begin{equation}
{\tilde{q}}^l_{\vec{r}_j}= {\tilde{s}}^l_{\vec{r}_j,s} + {\tilde{s}}^l_{\vec{r}_j,\eta} \, ,
\hspace{0.25cm} l=\pm,z \, .
\label{q-oper}
\end{equation}

We emphasize that the $c$ fermion operators, Eq. (\ref{fc+}), and the spinon and $\eta$-spinon operators
defined by Eqs.  (\ref{Scs-j-rot}), (\ref{rotated-quasi-spin}), and (\ref{sir-pir}) are mapped from the rotated-electron 
operators by an exact local unitary transformation that does not introduce constraints. Given their direct 
relation to the generators of the model's extended global symmetry, their occupancy configurations naturally 
generate representations of the corresponding global symmetry algebra. Consistent with the lack of constraints of such a 
local unitary transformation, inversion of the relations given in Eqs. (\ref{fc+}) and (\ref{rotated-quasi-spin}) 
fully defines the rotated-electron operators in terms of the $c$ fermion and $\eta s$ quasi-spin
operators as follows,
\begin{eqnarray}
{\tilde{c}}_{\vec{r}_j,\uparrow}^{\dag} & = &
f_{\vec{r}_j,c}^{\dag} \left({1\over 2} - {\tilde{q}}^{z}_{\vec{r}_j}\right) + e^{i\vec{\pi}\cdot\vec{r}_j}\, f_{\vec{r}_j,c}
\left({1\over 2} +{\tilde{q}}^{z}_{\vec{r}_j}\right) \, ,
\nonumber \\
{\tilde{c}}_{\vec{r}_j,\downarrow}^{\dag} & = &
(f_{\vec{r}_j,c}^{\dag} 
+ e^{i\vec{\pi}\cdot\vec{r}_j}\,f_{\vec{r}_j,c})\,{\tilde{q}}^+_{\vec{r}_j} \, ,
\nonumber \\
{\tilde{c}}_{\vec{r}_j,\uparrow} & = &  f_{\vec{r}_j,c}\left({1\over 2} 
-{\tilde{q}}^{z}_{\vec{r}_j}\right) + e^{i\vec{\pi}\cdot\vec{r}_j}\,f_{\vec{r}_j,c}^{\dag}
\left({1\over 2} + {\tilde{q}}^{z}_{\vec{r}_j}\right)  \, ,
\nonumber \\
{\tilde{c}}_{\vec{r}_j,\downarrow} & = &
(f_{\vec{r}_j,c} +e^{i\vec{\pi}\cdot\vec{r}_j}\,f_{\vec{r}_j,c}^{\dag})\,{\tilde{q}}^-_{\vec{r}_j} \, .
\label{c-up-c-down}
\end{eqnarray}
As given in Eq. (\ref{albegra-cf-s-h}) of Appendix \ref{commute} that the $c$ fermion operators 
commute with the $\eta s$ quasi-spin operators is behind the form of the expressions given here, whose 
$c$ fermion creation and annihilation operators are located on the left-hand side.

The $c$ fermion operator and $\eta s$ quasi-spin operator expressions in terms 
of rotated-electron creation and annihilation operators given in Eqs. (\ref{fc+})
and (\ref{rotated-quasi-spin}), respectively, are except for unimportant phase factors
similar to those considered in the studies of Refs. \cite{Ageluci,Mura-03,Stellan-06}
in terms of electron creation and annihilation operators. Our operational representation
has the advantage of rotated-electron single and double occupancy being good
quantum numbers for all finite interaction values. 
On the other hand, the operator expressions provided in Eqs. (\ref{fc+})
and (\ref{rotated-quasi-spin}) differ from those of Refs. \cite{companion,companion0}
by unimportant phase factors.

Since for finite $U/t$ values the Hamiltonian $\hat{H}$ of Eq. (\ref{HH}) does not commute with 
the unitary operator ${\hat{V}} = e^{-{\hat{S}}}$, when expressed in terms 
of the rotated-electron creation and annihilation operators of Eq. (\ref{rotated-operators}) it has 
an infinite number of terms, 
\begin{eqnarray}
{\hat{H}} & = & {\hat{V}}\,{\tilde{H}}\,{\hat{V}}^{\dag}
= {\tilde{H}} + [{\tilde{H}},{\tilde{S}}\,] 
\nonumber \\
& + & {1\over 2}\,[[{\tilde{H}},{\tilde{S}}\,],{\tilde{S}}\,] + ... \, .
\label{HHr}
\end{eqnarray}
The commutator $[{\tilde{H}},{\tilde{S}}\,]$ does not vanish
except for $U/t\rightarrow\infty$ so that ${\hat{H}} \neq {\tilde{H}}$ for finite values of $U/t$. 

Provided that both $U/t$ is finite and one accounts for 
all higher-order terms on the right-hand-side of Eq. (\ref{HHr}), the corresponding
expression refers to the Hubbard model. This is in contrast to the physical problem
studied in Refs. \cite{Stein,HO-04,Mac,Harris}, for which 
the rotated creation and annihilation operators of Eq. (\ref{rotated-operators}) refer to
electrons. Thus except for $U/t\rightarrow\infty$ within the physical problem
studied in Refs. \cite{Stein,HO-04,Mac,Harris} the Hamiltonian given
in Eq. (\ref{HHr}) is not the Hubbard Hamiltonian. Instead, it is a rotated Hamiltonian
for which electron double occupancy and single occupancy are good quantum numbers.
On the other hand, for the alternative physical problem studied here and in Refs. \cite{companion,companion0}
the rotated creation and annihilation operators of Eq. (\ref{rotated-operators}) refer to
rotated electrons and the Hamiltonian provided in Eq. (\ref{HHr}) is the Hubbard Hamiltonian.

The latter Hamiltonian may be developed into an expansion  
whose terms can be written as products of the rotated kinetic operators ${\tilde{T}}_{\gamma}$
given in Eq. (\ref{T-op-rot}) of Appendix \ref{commute} where $\gamma =0,\pm 1$.
The corresponding order of a given Hamiltonian term refers to the number of such
rotated kinetic operators ${\tilde{T}}_{\gamma}$ independently of
their type, $\gamma =0,\pm 1$. To fourth order such an Hamiltonian reads,
\begin{eqnarray}
{\hat{H}} & = & {\hat{H}}^{(0)} + {\hat{H}}^{(1)} +  {\hat{H}}^{(2)} +
{\hat{H}}^{(3)} +  {\hat{H}}^{(4)} + ... \, ,
\nonumber \\
{\hat{H}}^{(0)} & = & U\,\tilde{V}_D 
\, ; \hspace{0.5cm} {\hat{H}}^{(1)} = t\,\tilde{T}_{0} \, ,
\nonumber \\
{\hat{H}}^{(2)} & = & - {t^2\over U}\,\tilde{T}_{-1}\tilde{T}_{+1} \, ,
\nonumber \\
{\hat{H}}^{(3)} & = & {t^3\over U^2}\,[\tilde{T}_{-1}\tilde{T}_{0}\tilde{T}_{+1}
- {1\over 2}(\tilde{T}_{-1}\tilde{T}_{+1}\tilde{T}_{0}
+ \tilde{T}_{0}\tilde{T}_{-1}\tilde{T}_{+1})]
\nonumber \\
{\hat{H}}^{(4)}  & = & {t^4\over U^3}\,[\tilde{T}_{-1}\tilde{T}_{0}\tilde{T}_{+1}\tilde{T}_{0} + \tilde{T}_{0}\tilde{T}_{-1}\tilde{T}_{0}\tilde{T}_{+1}
\nonumber \\
& - & \tilde{T}_{-1}\tilde{T}_{0}^2\tilde{T}_{+1} -  {1\over 2}\tilde{T}_{-1}^2\tilde{T}_{+1}^2
\nonumber \\
& + & \tilde{T}_{-1}\tilde{T}_{+1}\tilde{T}_{-1}\tilde{T}_{+1}
- {1\over 2}(\tilde{T}_{-1}\tilde{T}_{+1}\tilde{T}_{0}^2
+ \tilde{T}_{0}^2\tilde{T}_{-1}\tilde{T}_{+1})
\nonumber \\
& + & \theta\,(2\tilde{T}_{0}\tilde{T}_{-1}\tilde{T}_{+1}\tilde{T}_{0}
- \tilde{T}_{-1}\tilde{T}_{+1}\tilde{T}_{0}^2 - \tilde{T}_{0}^2\tilde{T}_{-1}\tilde{T}_{+1})] \, ,
\nonumber \\
\theta  & - & \hspace{0.15cm}{\rm real-number}\hspace{0.15cm}{\rm parameter}   \, .
\label{HHr-domi}
\end{eqnarray}
Here,
\begin{eqnarray}
{\tilde{V}}_D & = & {\hat{V}}^{\dag}\,{\hat{V}}_D\,{\hat{V}}
\nonumber \\
& = & \sum_{j=1}^{N_a}\left(\tilde{n}_{\vec{r}_j,\uparrow} -1/2\right)
\left(\tilde{n}_{\vec{r}_j,\downarrow}-1/2\right) \, .
\label{tilde-VD}
\end{eqnarray}
is the rotated-electron interaction operator. That it appears only once in the Hamiltonian 
expansion whose leading-order terms are given in Eq. (\ref{HHr-domi}) follows from the derivation of that expansion 
systematically using the commutator,
\begin{equation}
[\tilde{V}_D,{\tilde{T}}_{\gamma}] = \gamma\,{\tilde{T}}_{\gamma} \, , \hspace{0.35cm} \gamma =0,\pm 1 \, .
\label{comm-V-T}
\end{equation}

We recall that except for $U/t\rightarrow\infty$ one has that $\hat{T}_0\neq \tilde{T}_0$, 
$\hat{T}_{+1}\neq \tilde{T}_{+1}$, and $\hat{T}_{-1}\neq \tilde{T}_{-1}$. Expressing the
Hamiltonina expression of Eq. (\ref{HHr-domi}) in terms of electron creation and
annihilation operators gives for large $U/t$ values a similar expansion. However, 
for the intermediate $U/t$ values of interest for our study the few first terms of the Hamiltonian $t/U$
expansion given in of Eq. (\ref{HHr-domi}) in terms of rotated-electron operators contain much more
complicated higher-order terms when expressed in terms of electron creation and
annihilation operators.

That Hamiltonian expansion may be expressed in terms 
of the $c$ fermion and $\eta s$ quasi-spin operators. This is achieved by combining the rotated-electron
operator expressions provided in Eq. (\ref{c-up-c-down}) with those of the rotated-electron interaction operator
given in Eq. (\ref{tilde-VD}) and three rotated 
kinetic operators ${\tilde{T}}_{0}$, ${\tilde{T}}_{-1}$, and ${\tilde{T}}_{+1}$ provided in
Eq. (\ref{T-op-rot}) of Appendix \ref{commute}.

If a rotated-electron term of an operator expansion in terms of rotated-electron 
creation and annihilation operators does not
preserve the numbers of rotated-electron singly and doubly occupied sites, we call it off-diagonal. An interesting 
technical detail is that up to third order all diagonal terms of the Hamiltonian 
expression provided in Eq. (\ref{HHr-domi}) are generated by the leading-order term of the 
operator ${\tilde{S}}$, which is given in Eq. (\ref{Sr-rot}). Indeed, when expressed 
in terms of electron operators the Hubbard Hamiltonian provided in Eq.
(\ref{HH}) does not contain any off-diagonal terms 
with more than two electron operators. (In this case the off-diagonal terms are electron off-diagonal 
terms, which refer to electron doubly occupied sites.)

Only the Hamiltonian terms ${\hat{H}}^{(0)}$, ${\hat{H}}^{(1)}$, ${\hat{H}}^{(2)}$, and ${\hat{H}}^{(3)}$
to third order given in Eq. (\ref{HHr-domi}) are universal. 
Indeed, the form of the terms of fourth and larger order is different for each electron - rotated-electron
unitary transformation. For the fourth-order term ${\hat{H}}^{(4)}$ given in that equation only
the real-number parameter $\theta$ value is not universal, being unitary-transformation dependent
\cite{HO-04}. For instance, the methods of Refs. \cite{Mac} and \cite{Harris}
refer to two different electron - rotated-electron unitary transformations whose
$\theta$ values are $\theta=0$ and $\theta =1/4$, respectively. Moreover,
one of the methods of Ref. \cite{HO-04} refers to an electron - rotated-electron unitary 
transformation whose $\theta$ value is $\theta=1/2$. Its value for the electron - rotated-electron 
unitary transformation whose unitary operator is uniquely defined by the matrix elements 
of Eqs. (\ref{ME-GEN}) and (\ref{ME-Vll}) remains an open issue.
Fortunately, these Hamiltonian terms multiplying the parameter $\theta$
vanish at half filling so that this does not affect the ensuing section studies.

The non-universal Hamiltonian terms
are all reducible with respect to the subspaces with fixed values
of rotated-electron single and double occupancies. That is, they contain hopping processes 
that do not originate from excitation between these subspaces, 
$\tilde{T}_{0}\tilde{T}_{-1}\tilde{T}_{+1}\tilde{T}_{0}$, nor terminate once a 
rotated-electron or rotated-hole is returned to a subspace with larger singly-occupancy, for example, 
$\tilde{T}_{-1}\tilde{T}_{+1}\tilde{T}_{-1}\tilde{T}_{+1}$. All these processes can be 
viewed as arising from the specific transformation ${\hat{V}}^{\dag}\vert \Psi_{l_r,l_{\eta s},\infty}\rangle$
of the $U/t\rightarrow\infty$ energy eigenstates within the subspaces with fixed values
of rotated-electron single and double occupancies. Thus the infinite
electron - rotated-electron unitary transformations differ in the processes
within each subspace with fixed values of these occupancies.

\section{General $S_s=1$ spin spectrum within the spinon representation}
\label{spin-spectrum-incoh}

As discussed in Sec. \ref{Introduction}, the usual spin-wave theory does not describe  
the neutron scattering of LCO. Here we study the $S_s =1$ spin-triplet spectrum of
the half-filled Hubbard model on the square lattice by means of the spinon representation
that emerges from the above more general $c$ fermion and $\eta s$ quasi-spin operator formulation,
which is that suitable for the LCO intermediate interaction range $U/t\in (6,8)$. 
(In units of the bandwidth, $8t$, this gives $U/8t\in (0.75,1)$.)

For very large $U/t$ values the Hubbard model may be mapped onto
a spin-only problem whose spins are those of the electrons that
singly occupy sites. However, for intermediate $U/t$ values electron
single occupancy is not a good quantum number so that such a mapping
breaks down. On the other hand, the rotated electrons of our operator representation
have been constructed to make rotated-electron
single and double occupancy good quantum numbers for $U/t>0$.
This is why our spinons are well defined for the LCO intermediate interaction range $U/t\in (6,8)$.
Indeed they are the spins of the rotated electrons that singly
occupy sites. In the large-$U/t$ limit the rotated-electrons become electrons, so that
one recovers the known standard results.

Within our operator formulation, the Hubbard model in the vanishing rotated-electron 
doubly occupied sites number and unoccupied sites number subspace (VDU subspace) can be mapped onto
a spin-only problem for all $U/t$ finite values. In the VDU subspace the number of spin-$1/2$ spinons equals that
of rotated-electrons, electrons, and sites $N=N_a$. Since there are no rotated-electron doubly occupied
or unoccupied sites there are no $\eta$-spinons. Hence the number of $\eta$-spin $SU(2)$ symmetry
algebra representations vanishes and that symmetry does not play any role.
Furthermore, although there are $N=N_a$ $c$ fermions,
their $c$ momentum band associated with the operators of Eq. (\ref{fc+q})
is full. Hence the degrees of freedom associated with the $c$ fermion occupancy configurations
that generate the $c$ hidden $U(1)$ symmetry algebra representations 
are frozen and the Hubbard model in the subspace under consideration may be mapped onto a spin-only problem, as
confirmed below. 

For $U/t\rightarrow\infty$ the VDU subspace is the only one for finite excitation energy. For the 
finite-$U/t$ spin excitations that preserve the electron number
$N=N_a$ considered in the following, it is the only subspace within a finite excitation-energy window, $\omega\in (0,2\Delta_{MH})$.
Here $2\Delta_{MH}$ is the Mott-Hubbard gap. Below we calculate its $U/t$ dependence for
the LCO intermediate interaction range $U/t\in (6,8)$ by DMRG. Our goal is to check whether the relevant spin 
energy spectrum that emerges from our VDU subspace spin-only problem is indeed contained in the 
excitation-energy domain $\omega\in (0,2\Delta_{MH})$ for which it is valid.

\subsection{The energy range of our spin-only quantum problem}
\label{2Delta-spinon-only}

From the interplay of the model's symmetries with our operator formulation that makes these 
symmetries explicit, one straightforwardly confirms that the minimum energy for creation of one rotated-electron
doubly occupied site or one rotated-electron unoccupied site at fixed electron number $N=N_a$
onto the $n=1$ and $m=0$ ground state is indeed given by
the Mott-Hubbard gap, $2\Delta_{MH}$. Its magnitude is twice that of the single-particle gap,
$\Delta_{MH}$. In order to define the energy range of our study, here we calculate the 
Mott-Hubbard gap $2\Delta_{MH}$ dependence on $U/t$ for a domain containing
the LCO range $U/t\in (6,8)$.
\vspace{0.5cm}
\begin{figure}[hbt]
\begin{center}
\centerline{\includegraphics[width=6.00cm]{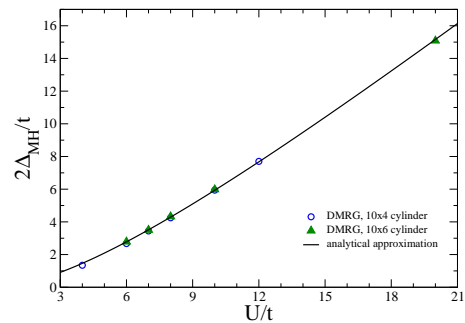}}
\caption{The Mott-Hubbard gap $2\Delta_{MH}$ DMRG numerical results on two different width cylinders
along with the approximate analytical expression of Eq. (\ref{ansatz}) (solid line) as a function of $U/t$.
The DMRG points seem to be consistent with for the half-filled Hubbard on the square lattice the Mott-Hubbard 
gap being finite for $U/t>0$ and vanishing in the $U/t\rightarrow 0$ limit.}
\label{fig3}
\end{center}
\end{figure}

Our DMRG calculations refer to the single-particle gap. They have
been performed both for $10\times4$ and $10\times6$ Hubbard cylinders. 
The chemical potential was set to $U/2$ and two states were targeted, one with $N$ particles and the other
with $N-1$. (Targetting $N+1$ electrons would have given the same results.) Both states were put into
the same density matrix in the traditional multi-state targeting DMRG approach. Thus, the
same truncation error applied to both states, leading to significant error cancellation. The
resulting gap at each sweep was plotted versus the maximum truncation error in the sweep,
yielding approximately linear behavior, and allowing the extrapolation to zero truncation
error. The error estimate is roughly the size of the extrapolation from the last point.
From $1800$ ($10\times 4$ Hubbard cylinder) to $6000$ ($10\times 6$ Hubbard cylinder) states were kept. 

Here we report the corresponding magnitudes of the Mott-Hubbard gap $2\Delta_{MH}$. 
For the range $U/t\in (4,20)$ we find that, 
\begin{equation}
2\Delta_{MH} \approx U\left[\tanh\left({\sqrt{U/\gamma\,t\over \sqrt{6}\,\gamma+\sqrt{U/\gamma\,t}}}\right)\right]^2 
\, ; \hspace{0.15cm}  \gamma = {\pi + 6\over 6} \, ,
\label{ansatz}
\end{equation}
gives quantitative agreement with our numerical DMRG calculations for the Mott-Hubbard gap $2\Delta_{MH}$ dependence on $U/t$. 
The DMRG points for that gap are plotted in Fig. \ref{fig3} along
with the curve obtained from the approximate analytical expression, Eq. (\ref{ansatz}). 

For instance, our DMRG calculations for $10\times 6$ Hubbard cylinders give
$2\Delta_{MH} \approx 2.78(4)\,t$ for $U/t=6$ and $\Delta_{MH} = 4.30(4)\,t$ for $U/t=8$. This leads to a range 
$2\Delta_{MH}\in (816$ meV$,1442$ meV) for $U/t\in (6,8)$. Here we used the $t$ magnitudes $t\approx 293$ meV 
and $t\approx 335$ meV for which the model describes the LCO neutron scattering for $U/t=6$ and $U/t=8$, respectively.
For the $U/t=6.1$ value used in some of our calculations, we find $\Delta_{MH} \approx 2.81(0)\,t$
from the DMRG analysis, so that $\Delta_{MH}\approx 829$ meV for $t\approx 295$ meV.

Optical experiments overestimate the charge-transfer gap magnitudes of the parent insulating compounds 
\cite{Moskvin-11}. On the other hand, by measuring the Hall coefficient $R_H$ in LCO, the studies
of Ref. \cite{Ono-07} have estimated the energy gap over which the electron and hole
carriers are thermally activated, which corresponds to the Mott-Hubbard gap,
to be $2\Delta_{MH}\approx 890$ meV. Remarkably, this magnitude is 
within the range $2\Delta_{MH}\in (816$ meV$,1442$ meV) of our above theoretical predictions for $U/t\in (6,8)$.
Our theoretical approach based on the combination of our DMRG results with
the $U$ and $t$ values for which agreement with the LCO neutron-scattering 
agreement is reached leads to $2\Delta_{MH}\approx 890$ meV for $U/t\approx 6.3$.
Below we consistently confirm that the spin-triplet excitation spectrum calculated for
the Hubbard model in the VDU subspace is contained in the
energy window $\omega\in (0,2\Delta_{MH})$ found here.

Note that the DMRG points of Fig. \ref{fig3} seem to be consistent with
the Mott-Hubbard gap being finite for $U/t>0$ and vanishing in the $U/t\rightarrow 0$ limit.

\subsection{The Hubbard model in the VDU subspace}
\label{HM-spinon-ope}

Let us confirm that within our operator representation
the half-filled Hubbard model on the square lattice in the VDU subspace can for $U/t>0$ be expressed solely in terms 
of spinon operators. Indeed, accounting for the lack of both rotated-electron doubly occupied sites
and unoccupied sites, upon writing the Hamiltonian of Eq. (\ref{HHr-domi}) in the VDU subspace, 
one finds that all its terms of odd order vanish and the terms of even order given
in that equation simplify to,
\begin{eqnarray}
{\hat{H}}^{(0)} & = & U\,\tilde{V}^c \, ,
\nonumber \\
{\hat{H}}^{(2)} & = & - {t^2\over U}\,\tilde{T}_{-1}\tilde{T}_{+1} \, ,
\nonumber \\
{\hat{H}}^{(4)}  & = & {t^4\over U^3}\,[\tilde{T}_{-1}\tilde{T}_{+1}\tilde{T}_{-1}\tilde{T}_{+1}
\nonumber \\
& - & {1\over 2}\tilde{T}_{-1}^2\tilde{T}_{+1}^2
-\tilde{T}_{-1}\tilde{T}_{0}^2\tilde{T}_{+1}] \, .
\label{HHr-domi-x0}
\end{eqnarray}

We have then expressed the Hamiltonian terms of even order as those provided
in Eq.  (\ref{HHr-domi-x0}) in terms 
of the $c$ fermion and $\eta s$ quasi-spin operators. This has been done by 
combining the rotated-electron operator expressions provided in Eq. (\ref{c-up-c-down}) with those of the three rotated 
kinetic operators ${\tilde{T}}_{0}$, ${\tilde{T}}_{-1}$, and ${\tilde{T}}_{+1}$ given in
Eq. (\ref{T-op-rot}) of Appendix \ref{commute}. Since the states that span
the VDU subspace are generated only by rotated-electron singly occupancy configurations, the projectors ${\tilde{q}}_{\vec{r}_j}^c$
and $(1-{\tilde{q}}_{\vec{r}_j}^c)$ in the expressions of Eq. (\ref{sir-pir}) can be replaced
by the corresponding eigenvalues $1$ and $0$, respectively. One then finds that
${\tilde{s}}^l_{\vec{r}_j,s} = {\tilde{q}}^l_{\vec{r}_j}$ in the VDU subspace, so that the $\eta$-spinon operators
do not play any role. Hence in it the $\eta s$ quasi-spin operators ${\tilde{q}}^l_{\vec{r}_j}$
reduce to the corresponding spinon operators ${\tilde{s}}^l_{\vec{r}_j,s}$, where $ l=\pm,z$.

Moreover, after some algebra involving the anti-commutation and commutation
relations given in Eqs. (\ref{albegra-cf})-(\ref{albegra-s-e-s-com}) of Appendix \ref{commute}
one finds that all contributions involving the
$c$ fermion creation and annihilation operators can be expressed 
only in terms of local operators ${\tilde{q}}^{c}_{j}$. In
the VDU subspace one can then replace these operators by their eigenvalue $1$. Thus
the Hamiltonian terms of Eq. (\ref{HHr-domi-x0}) can be expressed only in terms
of spinon operators. Importantly, this holds as well for the remaining Hamiltonian
terms of higher even order omitted in that equation. Moreover, all Hamiltonian terms
of odd order vanish and the zeroth-order term becomes a mere constant, 
${\hat{H}}^{(0)} = [U/4]\,N_a$, and may be ignored. The Hamiltinonian terms
of second and fourth order of Eq. (\ref{HHr-domi-x0}) may after some
algebra then be rewritten as,
\begin{equation}
{\hat{H}}^{(2)} = {t^2\over U}\sum_{\langle j_1 j_2\rangle} {1\over 2}[{\vec{\tilde{s}}}_{\vec{r}_{j_1},s}\cdot{\vec{\tilde{s}}}_{\vec{r}_{j_2},s}-1] \, ,
\label{HHr-spinons-2}
\end{equation}
and
\begin{eqnarray}
{\hat{H}}^{(4)} & = & - {t^4\over U^3}\sum_{\langle j_1 j_2\rangle} {1\over 2}[{\vec{\tilde{s}}}_{\vec{r}_{j_1},s}\cdot{\vec{\tilde{s}}}_{\vec{r}_{j_2},s}-1]
\nonumber \\
& + & {t^4\over U^3}\sum_{j_1,j_2,j_3} {1\over 2}D_{j_1,j_2} D_{j_2,j_3} [{\vec{\tilde{s}}}_{\vec{r}_{j_1},s}\cdot{\vec{\tilde{s}}}_{\vec{r}_{j_3},s}-1]
\nonumber \\
& + & {t^4\over U^3}\sum_{j_1,j_2,j_3,j_4} {1\over 8}D_{j_1,j_2} D_{j_2,j_3} D_{j_3,j_4}D_{j_4,j_1}
\nonumber \\
& \times & [1-{\vec{\tilde{s}}}_{\vec{r}_{j_1},s}\cdot{\vec{\tilde{s}}}_{\vec{r}_{j_2},s} - {\vec{\tilde{s}}}_{\vec{r}_{j_1},s}\cdot{\vec{\tilde{s}}}_{\vec{r}_{j_3},s}
- {\vec{\tilde{s}}}_{\vec{r}_{j_1},s}\cdot{\vec{\tilde{s}}}_{\vec{r}_{j_4},s}
\nonumber \\
& - & {\vec{\tilde{s}}}_{\vec{r}_{j_2},s}\cdot{\vec{\tilde{s}}}_{\vec{r}_{j_3},s}
- {\vec{\tilde{s}}}_{\vec{r}_{j_2},s}\cdot{\vec{\tilde{s}}}_{\vec{r}_{j_4},s}-{\vec{\tilde{s}}}_{\vec{r}_{j_3},s}\cdot{\vec{\tilde{s}}}_{\vec{r}_{j_4},s}]
\nonumber \\
& + & {t^4\over U^3}\sum_{j_1,j_2,j_3,j_4} {5\over 8}D_{j_1,j_2} D_{j_2,j_3} D_{j_3,j_4}D_{j_4,j_1}
\nonumber \\
& \times &
[({\vec{\tilde{s}}}_{\vec{r}_{j_1},s}\cdot{\vec{\tilde{s}}}_{\vec{r}_{j_2},s})({\vec{\tilde{s}}}_{\vec{r}_{j_3},s}\cdot{\vec{\tilde{s}}}_{\vec{r}_{j_4},s})
\nonumber \\
& + & ({\vec{\tilde{s}}}_{\vec{r}_{j_1},s}\cdot{\vec{\tilde{s}}}_{\vec{r}_{j_4},s})({\vec{\tilde{s}}}_{\vec{r}_{j_2},s}\cdot{\vec{\tilde{s}}}_{\vec{r}_{j_3},s})
\nonumber \\
& - & ({\vec{\tilde{s}}}_{\vec{r}_{j_1},s}\cdot{\vec{\tilde{s}}}_{\vec{r}_{j_3},s})({\vec{\tilde{s}}}_{\vec{r}_{j_2},s}\cdot{\vec{\tilde{s}}}_{\vec{r}_{j_4},s})] \, ,
\label{HHr-spinons-4}
\end{eqnarray}
respectively. Here the spinon operator ${\vec{\tilde{s}}}_{\vec{r}_j,s}$ has operator Cartesian 
components ${\tilde{s}}^{x}_{\vec{r}_j,s} = {1\over 2}[{\tilde{s}}^{+}_{\vec{r}_j,s} +{\tilde{s}}^{-}_{\vec{r}_j,s}]$, 
${\tilde{s}}^{y}_{\vec{r}_j,s} = {1\over 2i}[{\tilde{s}}^{+}_{\vec{r}_j,s} -{\tilde{s}}^{-}_{\vec{r}_j,s}]$,
and ${\tilde{s}}^{z}_{\vec{r}_j,s}$ and refers to the spin of a rotated electron that singly occupies
the site of real-space coordinate $\vec{r}_j$. The spinon operators ${\tilde{s}}_{\vec{r}_j,s}^{z}$ and ${\tilde{s}}^{\pm}_{\vec{r}_j,s}$ 
are those given in Eq. (\ref{Scs-j-rot}). Furthermore, in the expressions of Eqs. (\ref{HHr-spinons-2}) and (\ref{HHr-spinons-4})
the summation $\langle j_1 j_2\rangle$ runs over nearest-neighboring sites
and $D_{j,j'}=1$ for the real-space coordinates ${\vec{r}}_j$ and ${\vec{r}}_{j'}$
corresponding to nearest-neigboring sites and $D_{j,j'}=0$ otherwise. 

For very large $U/t$ values when
electron single and double occupancy become good quantum numbers 
and thus the rotated electrons become electrons the
spinon operators ${\vec{\tilde{s}}}_{\vec{r}_j,s}$ become the usual spin operators
${\vec{\hat{s}}}_{\vec{r}_j,s}$ and Eqs. (\ref{HHr-spinons-2}) and (\ref{HHr-spinons-4}) recover the corresponding 
spin-only Hamiltonian terms obtained previously by other authors \cite{Stein}.
On the other hand, for the intermediate $U/t$ values of interest for LCO the terms of the Hamiltonian $t/U$
expansion given in of Eq. (\ref{HHr-spinons-4}) in terms of spinon (rotated-electron) operators contain much more
complicated higher-order terms when expressed in terms of electron creation and
annihilation operators.

\subsection{The absolute ground state of the Hubbard model on the square lattice}
\label{GS}

The antiferromagnetic long-range order of the half-filled Hubbard model on
the square lattice ground state follows from a spontaneous
symmetry breaking mechanism that occurs in the thermodynamic
limit $N_a\rightarrow\infty$. It involves a whole tower of low-lying energy eigenstates of
the finite system. They collapse in that limit onto the ground state. 

Importantly, both that ground state and 
the excited energy eigenstates that collapse onto it as
$N_a\rightarrow\infty$ belong to the VDU subspace. 
One may investigate which energy eigenstates couple to the 
exact finite $N_a\gg 1$ and $n=1$ and $m=0$ ground state $\vert GS\rangle$ via the operator,
\begin{equation}
{\hat{m}}_s^{l} = {1\over N_a}\sum_{j=1}^{N_a} (-1)^j\,{\hat{s}}^{l}_{\vec{r}_j,s} \, ,
\hspace{0.35cm} l = \pm, z \, .
\label{Msl}
\end{equation}
We insert the complete set of energy eigenstates as follows,
\begin{eqnarray}
& & \langle GS\vert({\hat{m}}_s^{l})^2\vert GS\rangle =
\sum_{l_r,l_{\eta s}}\langle GS\vert{\hat{m}}_s^{l}\vert \Psi_{l_r,l_{\eta s},u}\rangle
\nonumber \\
& \times & \langle\Psi_{l_r,l_{\eta s},u}\vert{\hat{m}}_s^{l}\vert \Psi_{GS}\rangle 
\nonumber \\
& = & \sum_{l_r,l_{\eta s}}\vert\langle GS\vert{\hat{m}}_s^{l}\vert \Psi_{l_r,l_{\eta s},u}\rangle\vert^2 
\, ; \hspace{0.25cm} l = \pm , x_3 \, .
\label{stag-magn}
\end{eqnarray}
Only energy eigenstates $\vert \Psi_{l_r,l_{\eta s},u}\rangle$ with excitation 
momentum $\vec{k}=\vec{\pi}$ and quantum numbers $S_{\eta}=0$, $2S_c=N_a=N$,
$S_s= 1$, and $S_s^{z}=0,\pm 1$ contribute 
to the sum of Eq. (\ref{stag-magn}). We recall that the quantum numbers $S_{\eta}=0$
and $2S_c=N_a=N$ remain unchanged and thus are the same as those of
the ground state $\vert GS\rangle$. We denote by $\vert \Psi_{1T}\rangle$
the $S_s=1$, $S_{\eta}=0$, $2S_c=N_a=N$, and $\vec{k}=[\pi,\pi]$ lowest spin-triplet state
whose excitation energy behaves as $1/N_a$ for finite
$N_a\gg 1$. For the range $U/t>4$ of interest for our studies 
the contribution from this lowest spin triplet state 
is by far the largest. For instance, for the related spin-$1/2$ Heisenberg model on the 
square lattice the matrix-element square 
$\vert\langle GS\vert{\hat{m}}_s^{l}\vert \Psi_{1T}\rangle\vert^2$ exhausts 
the sum in Eq. (\ref{stag-magn}) by more than 98.7\% \cite{Peter-88}.
A similar behavior is expected for the Hubbard model on the square lattice,
at least provided that $U/t>4$.

The special properties with respect to the lattice symmetry group of the
lowest energy eigenstates contributing to the linear Goldstone modes of the
corresponding $S_s=1$ spin-wave spectrum reveal the space-symmetry breaking 
of the $N_a\rightarrow\infty$ ground state. In the present case 
of the half-filled Hubbard model on the square lattice the translation symmetry is broken. 
Hence as found here both the $\vec{k}=[0,0]$ 
and $\vec{k}=[\pi,\pi]$ excitation momenta appear among the
lowest energy eigenstates contributing to the linear Goldstone modes of the
$S_s=1$ spin-wave spectrum. However, that the transitions to the
lowest spin-triplet state $\vert \Psi_{1T}\rangle$ of momentum $\vec{k}=[\pi,\pi]$
nearly exhaust the sum in Eq. (\ref{stag-magn}) is consistent with the first-moment 
sum rules of an isotropic antiferromagnet, such that no weight is generated by states 
of momentum $\vec{k}=[0,0]$.

One of the few exact theorems that apply to the half-filled Hubbard model on a
bipartite lattice and thus on a square lattice is that for a finite number of lattice
sites $N_a$ its ground state is a spin-singlet state \cite{Lieb89}. The studies
of Refs. \cite{companion,companion0} use an operator representation that
differs from that used here only by unimportant phase factors. Such
studies provide evidence that the $n=1$ and $m=0$ ground state is the only
model's ground state that is invariant under the electron - rotated-electron unitary
transformation. For $N_a\gg 1$ the results of those references reveal that
its spin-singlet configurations refer to $N_a/2=N/2$ independent spin-singlet two-spinon 
pairs. Most of the weight of these spin-singlet two-spinon 
pairs stems from spinons at nearest-neighboring sites yet
they have finite contributions as well from spinons located at larger distances.  

Our spinon representation has been constructed to make
such $N/2$ spin-singlet spinon pairs correspond to spin-neutral objects that obey a hard-core
bosonic algebra. One can then perform an extended Jordan-Wigner
transformation that maps them onto $N/2$ $s1$ fermions \cite{companion0}. (In the index
$s1$ the number $1$ refers to one spin-singlet spinon pair.)
The corresponding $s1$ fermion momentum band is full for
the $n=1$ and $m=0$ absolute ground state. It has a momentum area $2\pi^2$ 
and coincides with an antiferromagnetic RBZ whose
momentum $\vec{q}$ components obey the inequality,
\begin{equation}
\vert q_{x}\vert+\vert q_{y}\vert\leq\pi \, .
\label{s1-band}
\end{equation}

As a result of its invariance under the electron - rotated-electron unitary
transformation, the $n=1$ and $m=0$ absolute ground state is the only ground state that for $U/t>0$
belongs to a single $V$ tower. Hence both for it and its spin-triplet excited states that
belong to the VDU subspace the $s1$ boundary-line momenta 
${\vec{q}}_{Bs1}$ are independent of $U/4t$. Consistent with Eq. (\ref{s1-band}), 
their Cartesian components $q_{Bs1x}$ and $q_{Bs1y}$ obey the equations,
\begin{eqnarray}
& & q_{Bs1x}\pm q_{Bs1y}=\pi \, ,
\nonumber \\
& {\rm or} &
\hspace{0.10cm} q_{Bs1x}\pm q_{Bs1y}=-\pi \, .
\label{g-FS-x-0}
\end{eqnarray}
Hence the $s1$ boundary line refers
to the lines connecting $[\pm\pi,0]$ and $[0,\pm\pi]$. 

\subsection{The spin excitations and the ground-state spinon $d$-wave pairing}
\label{spin-ex-d-wave}

Within our spinon operator representation the $S_s=1$ spin-triplet excitations relative to 
the $n=1$ and $m=0$ absolute ground state involve creation of two holes in the $s1$ band 
along with a shift $\vec{\pi}/N_a$ of all discrete momentum values of the full $c$ band.
Under such an excitation one of the $N_a/2=N/2$ spin-singlet spinon pairs is broken.
This gives rise to two unbound spinons in the excited state whose three occupancy
configurations generate the three spin-triplet states of spin projection $S_s^{z}=0,\pm 1$.
In the case of such spin-triplet excitations the occupancy configurations of the two 
holes arising in the $s1$ fermion momentum band may simulate the motion of the
two unbound spinons relative to a background of $N/2-1$ spinon pairs, or vice versa.

The general spin-triplet spectrum has within the present spinon representation
the following form, 
\begin{eqnarray}
\omega (\vec{k}) & = & -\epsilon_{s1} ({\vec{q}}) -\epsilon_{s1} ({\vec{q}}\,') \, ,
\nonumber \\
\vec{k} & = & \vec{\pi} - {\vec{q}} - {\vec{q}}\,' \, ,
\label{DE-spin-x0}
\end{eqnarray}
where $\vec{\pi}=\pm [\pi,\pm\pi]$, ${\vec{q}}$ and ${\vec{q}}\,'$ are the momentum
values of the emerging two $s1$ fermion holes, and $\epsilon_{s1} ({\vec{q}})$ is the
corresponding $s1$ fermion energy dispersion. Indeed, the results of Ref. \cite{companion0}
provide evidence that for the Hubbard model on the square lattice in the
VDU subspace the $s1$ fermion momentum ${\vec{q}}$ is a good quantum number,
so that one can define a corresponding energy dispersion. However, in contrast
to 1D this property does not hold for the more general problem of that
model in its full Hilbert space \cite{companion0}.

The Hubbard model on the also bipartite 1D lattice has the same extended global symmetry
than on the square lattice. Hence for it an operator representation similar to that used here may 
be introduced. The exact Bethe-anstaz solution then implicitly performs the summation of all Hamiltonian 
terms of even order whose leading-order terms are given in Eqs. (\ref{HHr-spinons-2}) and (\ref{HHr-spinons-4}). This leads to a
$s1$ fermion band $\epsilon_{s1} (q)$ that in the $U/t\rightarrow 0$ limit equals the occupied
part of the electron non-interacting dispersion \cite{1D-03,1D-04}. The main effect of increasing the $U/t$ value
is decreasing the $s1$ fermion band $\epsilon_{s1} (q)$ energy bandwidth. It decreases
from $2t$ as $U/t\rightarrow 0$ to zero for $U/t\rightarrow\infty$.
\begin{figure}[hbt]
\begin{center}
\centerline{\includegraphics[width=6.5cm]{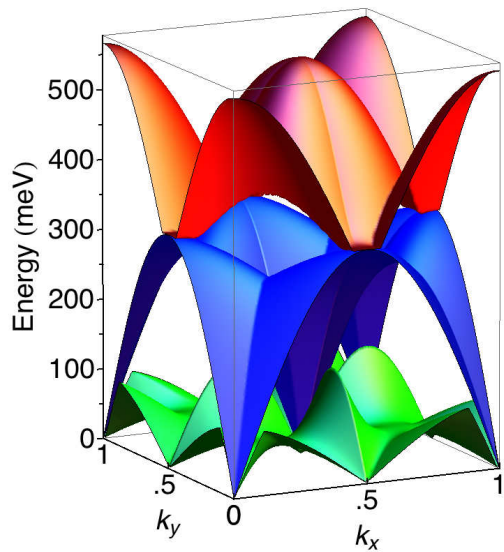}}
\caption{The energy-momentum space limits of the spin $S_s =1$ excited states spectrum of Eq. (\ref{DE-spin-x0}) for $U/t=6.1$, 
$t=295$\,meV, and $k_x$ and $k_y$ in units of $2\pi$. States whose energy is for a given $\vec{k}$ lower than
that of the intermediate spin-wave sheet as well as those of any energy and equivalent momenta 
$[0,0]=[0,2\pi]=[2\pi,0]=[2\pi,2\pi]$ do not contribute to the spin spectral weight.}
\label{fig4}
\end{center}
\end{figure}

As discussed above, expression of the Hamiltonian in terms of rotated-electron
operators leads for the intermediate $U/t\in (6,8)$ range to a quantum problem in terms
of rotated-electron processes similar to the corresponding large-$U/t$ quantum 
problem in terms of electron processes. At half filling the main effect of decreasing $U/t$ is
the increase of the energy bandwidth of an effective band associated with the spinon
occupancy configurations. Such an effective band is the $s1$ energy dispersion. Consistent with
and partially motivated by the exact 1D results yet accounting both for the corresponding common global symmetry 
and different physics, the rotated-electron studies of Refs. \cite{companion,companion0} provide evidence
that for the model on the square lattice the effective $s1$ energy dispersion $\epsilon_{s1} ({\vec{q}})$
involves an auxiliary dispersion,
\begin{equation}
\epsilon^0_{s1} (\vec{q}) = - {W^0_{s1}\over 2}[\cos q_{x} +\cos q_{y}] \, .
\label{Esx0}
\end{equation}
In the $U/t\rightarrow 0$ limit such an auxiliary dispersion reaches its maximum energy
bandwidth. Similarly to 1D, in that limit it is expected to become the occupied part of the electron non-interacting 
dispersion. The main effect of increasing $U/t$ is to decrease the energy bandwidth of that
dispersion and thus the magnitude of the energy
scale $W^0_{s1}$ in Eq. (\ref{Esx0}), so that for half filling it changes from $W^0_{s1}=4t$ as $U/t\rightarrow 0$ 
to $W^0_{s1}=0$ for $U/t\rightarrow\infty$.

However, the $n=1$ and $m=0$ ground state of the 1D half-filled Hubbard model has no antiferromagnetic 
long range order as $N_a\rightarrow\infty$. In the presence of that order, provided that
$U/t$ is not too small so that one can ignore the amplitude fluctuations of the corresponding order 
parameter, the problem can be handled for the model on the square lattice by a suitable mean-field theory. 
Within it the occurrence of that order is described by a $s1$ energy dispersion of 
the general form \cite{companion,companion0},
\begin{equation}
\epsilon_{s1} ({\vec{q}}) = -\sqrt{\vert\epsilon^0_{s1} ({\vec{q}})\vert^2 +\vert\Delta_{s1} ({\vec{q}})\vert^2} \, .
\label{bands}
\end{equation}
Here $\epsilon^0_{s1} ({\vec{q}})$ is the auxiliary dispersion given in Eq. (\ref{bands}) and
the gap function $\vert\Delta_{s1} ({\vec{q}})\vert$ is to be determined
from comparison with the spin-triplet spectrum obtained from the standard formalism of 
many-body physics by summing up an infinite number of ladder diagrams. (We
note that as explicitly shown in Ref. \cite{peres2002} the RPA studies of Sec. \ref{spin-spectrum-cohe} 
are equivalent to summing up an infinite number of such diagrams.)

We profit from symmetry and limit our analysis of the spin spectrum of Eq. (\ref{DE-spin-x0})
to the sector $k_x\in (0,\pi)$ and $k_y\in (0,k_x)$ of the $(\vec{k},\omega)$ space. Surprisingly, quantitative agreement with the results 
obtained from summing up an infinite number of diagrams is reached provided that the $s1$ dispersion gap 
function refers to a $d$-wave $s1$ fermion spin-singlet spinon pairing,
\begin{eqnarray}
\vert\Delta_{s1} (\vec{q})\vert = {\mu^0\over 2}\,{\vert \cos q_{x} -\cos q_{y}\vert
\over 2} \, .
\label{Delta-s1-q}
\end{eqnarray}
Moreover, from comparison with many-body physics results one finds that the inelastic coherent spin-wave
spectrum is generated by processes where $\vec{q}$ points in the nodal direction and $\vec{q}\,'$
belongs to the boundary of the $s1$ band reduced zone. The remaining
choices of $\vec{q}$ and $\vec{q}\,'$ either generate the inelastic incoherent continuum spectral weight
or vanishing weight, respectively. 

For this choice of the momenta of the two emerging $s1$ fermion holes
one finds from the use of Eqs. (\ref{Esx0})-(\ref{Delta-s1-q}) that the spin-wave spectrum 
corresponds to a surface of energy and momentum given by,
\begin{eqnarray}
\omega (\vec{k}) & = &
{\mu^0\over 2}\left\vert\sin\left({k_x+k_y\over 2}\right)\right\vert
+ W^0_{s1}\left\vert\sin\left({k_x-k_y\over 2}\right)\right\vert \, ,
\nonumber \\
\vec{k} & = & \vec{\pi} - {\vec{q}} - {\vec{q}}\,'  \, .
\label{om-SW}
\end{eqnarray} 
This is a particular case of the general spin spectrum of Eq. (\ref{DE-spin-x0}), which refers to the following choices of the momenta $\vec{\pi}$, ${\vec{q}}$, 
and ${\vec{q}}\,'$,
\begin{eqnarray}
\vec{\pi} & = & [\pi,-\pi] \, ,
\nonumber \\
{\vec{q}} & = & \left[{\pi\over 2}-{(k_x+k_y)\over 2},-{\pi\over 2}-{(k_x+k_y)\over 2}\right] \, ,
\nonumber \\
{\vec{q}}\,' & = & \left[{\pi\over 2}-{(k_x-k_y)\over 2},-{\pi\over 2}+{(k_x-k_y)\over 2}\right] \, ,
\label{k-SW-I}
\end{eqnarray} 
for the sub-sector such that $k_x\in (0,\pi)$, $k_y\in (0,k_x)$ for $k_x\leq\pi/2$,
and $k_y\in (0,\pi-k_x)$ for $k_x\geq\pi/2$. Moreover, for the sub-sector such that $k_y\in (0,\pi)$, $k_x\in (\pi-k_y,\pi)$ for $k_y\leq\pi/2$,
and $k_x\in (k_y,\pi)$ for $k_y\geq\pi/2$, respectively, it corresponds to the following choices of the momenta $\vec{\pi}$, ${\vec{q}}$, 
and ${\vec{q}}\,'$,
\begin{eqnarray}
\vec{\pi} & = & [\pi,\pi] \, ,
\nonumber \\
{\vec{q}} & = & \left[{\pi\over 2}-{(k_x+k_y)\over 2},{3\pi\over 2}-{(k_x+k_y)\over 2}\right] \, ,
\nonumber \\
{\vec{q}}\,' & = & \left[{\pi\over 2}-{(k_x-k_y)\over 2},-{\pi\over 2}+{(k_x-k_y)\over 2}\right] \, .
\label{k-SW-II}
\end{eqnarray} 
Note that the components of the $s1$ band momenta ${\vec{q}}$ appearing 
in Eqs. (\ref{k-SW-I}) and (\ref{k-SW-II}) are such that
$q_{x}-q_{y}=-\pi$ and thus belong to the half-filling $s1$ boundary line defined by Eq. (\ref{g-FS-x-0}), whereas 
those of the momenta ${\vec{q}}\,'$ in the same equations obey the relation
${q'}_{x}=-{q'}_{y}$ so that point in the nodal directions. 
\begin{figure}[hbt]
\includegraphics[width=6cm,height=6cm]{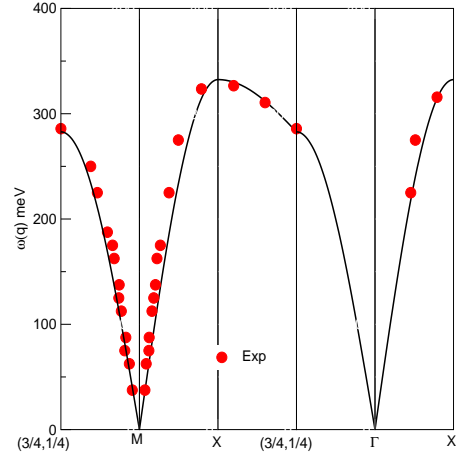}
\caption{The theoretical spin spectra, Eqs. (\ref{direc-GO})-(\ref{direc-XO}), 
(solid lines) plotted in the second BZ for $U/t\approx 6.1$
and $t\approx 0.295$ eV and thus
$\mu^0= 565.6$ meV and $W^0_{s1}= 49.6$ meV and the 
experimental data of Ref. \cite{LCO-2001} (circles) in meV. The 
momentum is given in units of $2\pi$. From Ref. \cite{companion}.}
\label{fig5}
\end{figure}

For the values $U/t=6.1$ and $t=295$ meV used in Sec. \ref{spin-spectrum-cohe}
in our study of the LCO spin spectrum one finds that $W^0_{s1}\approx t/5.95$ 
and $\mu^0\approx t/0.5216$ in the expressions of Eqs. (\ref{Esx0}), (\ref{Delta-s1-q}),
and (\ref{om-SW}), so that $W^0_{s1}\approx 49.6$ meV and $\mu^0\approx 565.6$ meV. 
The spin-wave spectrum of Eq. (\ref{om-SW}) calculated for these $W^0_{s1}$ and $\mu^0$ 
values refers to the middle surface plotted in Fig.
\ref{fig4}. Its expressions corresponding to the high symmetry directions in the BZ
are given in Appendix \ref{SS-spectrum}. The corresponding curves are plotted in the top
panel of Fig. \ref{fig2}, along with those obtained from the many-body physics by summing up 
an infinite number of ladder diagrams and the LCO experimental points of Ref. \cite{headings2010}.
In Fig. \ref{fig5} the same curves are plotted together with 
the LCO experimental points of Ref. \cite{LCO-2001}. We emphasize that the intermediate sheet 
plotted in Fig.\ref{fig4}, which corresponds to the general spin-wave spectrum of Eq. (\ref{om-SW}),
also fully agrees with the results of the experimental studies reported in Refs. \cite{headings2010,LCO-2001}.

The studies of Ref. \cite{companion} are limited to the spin-wave spectrum. Following the agreement
of the spin-wave spectrum of Eq. (\ref{om-SW}) obtained from the general spin spectrum of Eq. (\ref{DE-spin-x0}) 
with both results from the many-body physics and LCO neutron-scattering experimental points of
Refs. \cite{headings2010,LCO-2001} here we consider it for all choices of the $s1$ fermion
hole momenta $\vec{q}$ and $\vec{q}\,'$. The corresponding energy-momentum space domain of the
spin $S_s=1$ excited states whose spectrum is provided in Eq. (\ref{DE-spin-x0}) is represented in Fig. \ref{fig4}
for $U/t=6.1$ and $t=295$ meV. A similar spectrum is obtained for the values $U/t=8.0$ and $t=335$\,meV 
of Ref. \cite{lorenzana2005}. 

The largest energy of the general spin spectrum of the half-filled Hubbard model
on the square lattice in the VDU subspace represented in Fig. \ref{fig4} is $566$ meV.
Hence the whole spin spectrum of Eq. (\ref{DE-spin-x0}) represented in that figure is contained
in the energy window $\omega\in (0,2\Delta_{MH})$ of the corresponding VDU subspace.
Indeed, from combination of the results of our DMRG calculations with the $t$ magnitudes
that lead to agreement with the spin-wave spectrum of LCO we have found that
$2\Delta_{MH}\in (816$ meV$,1442$ meV) for $U/t\in (6,8)$.

As mentioned above, the intermediate sheet of the general spin spectrum represented in Fig. \ref{fig4}
refers to the spin-wave spectrum. For each excitation momentum $\vec{k}$, states of energy lower than the latter spectrum do not contribute 
to the form-factor weight. Furthermore and consistent with the first-moment sum rules of an isotropic antiferromagnet, 
no and nearly no weight is generated by states of any energy and momentum $[0,0]=[0,2\pi]=[2\pi,0]=[2\pi,2\pi]$ and near it, respectively.

Unfortunately, in its present form our spinon-operator method does not provide
the detailed continuum weight intensity distribution. However, it is expected that, similarly to the Heisenberg model case \cite{lorenzana2005}, 
its energy-integrated intensity follows the same trend as the spin-wave intensity. 
Analysis of Fig. \ref{fig4} reveals that for momentum $[\pi,0]$ there are no excited states of energy higher than
the spin waves. Thus at momentum $[\pi,0]$, the continuum weight distribution energy-integrated 
intensity vanishes or is extremely small, due to $s1$ band four-hole processes. Given the expected common trend of both intensities,
this is consistent with a damping of the spin-wave intensity at momentum $[\pi,0]$, as observed in the recent
high-energy inelastic neutron scattering experiments of Ref. \cite{headings2010}
but not captured by the Fig. \ref{fig2} RPA intensity. Since we find that for the half-filled
Hubbard model on the square lattice the continuum weight distribution energy-integrated 
intensity vanishes or is extremely small at momentum $[\pi,0]$, we predict that for it
the corresponding spin-wave intensity at momentum $[\pi,0]$ is also damped. Hence
we expect that such a behavior is absent in Fig. \ref{fig2} due to the RPA used
in the calculation of the spin-wave intensity. However, for all other momentum values
the RPA results are expected to be a quite good estimate of the model's spin-wave intensity. 

The $d$-wave spin-singlet spinon pairing that follows from the energy gap Eq. (\ref{Delta-s1-q}) emerged
here from imposing quantitative agreement with the spin-wave spectrum obtained from summing up an infinite 
number of diagrams. We emphasize that such a type of spinon pairing is not inconsistent with the ground-state 
antiferromagnetic order provided that the weights of the corresponding spinon pairs fall off as a power law of 
the spinon distance whose negative exponent has absolute value smaller than $5$. This was confirmed in
Ref. \cite{old-88} for spins associated with electrons yet holds as well for the present spinons,
which refer to the spins of the rotated-electrons that singly occupy sites.
We emphasize that such a pairing does not refer to electrons or rotated electrons. For the 
$n=1$ and $m=0$ absolute ground state it corresponds to $N/2$ spinon pairs, which describe the
spin degrees of freedom of the $N$ rotated electrons. Indeed, for that ground state all
rotated electrons singly occupy sites.

\section{Comparison of the predicted spectral weights with those in the LCO high-energy neutron scattering}
\label{LCOspin-spectrum}

As discussed in Sec. \ref{spin-spectrum-cohe}, the total spin-weight sum-rule, $\mu_B^2\,2(1-2d)$,
of Eq. (\ref{sr-eff}) can in units of $\mu_B^2$ be written as,
\begin{equation}
W_T = 2(1-2d) \approx W_{SW} + W_{CO} + 4(m_{AF}^{GB})^2 \, .
\label{WT}
\end{equation}
Here $W_{SW} = Z_d\,[2(1-2d)-4(m_{AF})^2]$ is the integrated spectral weight associated with the spin-wave
intensity, Eqs. (\ref{W-SW}) and (\ref{W-SW-exp}), $W_{CO} = (1-Z_d)\,[2(1-2d)-4(m_{AF})^2]$ that of
the remaining inelastic spin spectral weight associated with the continuum distribution,
and $4(m_{AF}^{GB})^2$ refers to the Bragg-peak elastic part of the spin spectral weight.

In Table \ref{table2} we provide the results of our calculations for several integrated spin spectral weights (in units of $\mu_B^2$).
This includes the total spin spectral weight $W_T$ and the spin-wave intensity coherent spectral weight 
$W_{SW}$. In addition, in the table we provide the estimated magnitudes of the total spin spectral weight
for excitation energy $\hbar\omega\leq$450 meV, $W_{<450}=W_{SW}/0.71 +4(m_{AF}^{GB})^2$, and
the total spin spectral weight for excitation energy $\hbar\omega >$ 450 meV, $W_{>450}=[W_T-W_{<450}]$.
The magnitude of the spin spectral weight for excitation energy $\hbar\omega\leq$450 meV, $W_{<450}$, 
is derived by combining our theoretical expressions with the 
observations of Ref. \cite{headings2010} that for the energy range up to about 450 meV,
71\% and 29\% of the weight corresponding to the inelastic response comes from the
coherent spin-wave weight and incoherent continuum weight, respectively.

Our prediction for the magnitude of the spin spectral weight for excitation energy $\hbar\omega\leq$450 meV,
$W_{<450}$, varies between $1.6\,\mu_B^2$ for $U/t\approx 6.1$ and $1.7\,\mu_B^2$ for $U/t\approx 8.0$,
in agreement with the experimental value $1.9\pm 0.3\,\mu_B^2$ reported in
Ref. \cite{headings2010}. 
\begin{table}
\begin{tabular}{|c|c|c|c|c|c|} 
\hline
$U/t$ & $6.1$ & $6.5$ & $8.0$ & $10.0$ \\
\hline
$W_T$ & $1.643$ & $1.671$ & $1.762$ ($1.778$ \cite{lorenzana2005}) & $1.848$ ($1.846$ \cite{lorenzana2005}) \\
\hline
$W_{SW}$ & $0.808$ & $0.799$ & $0.761$ & $0.714$ \\
\hline
$W_{<450}$ & $1.571$ & $1.593$ & $1.663$ & $1.730$ \\
\hline
$W_{>450}$ & $0.072$ & $0.078$ & $0.099$ & $0.118$ \\
\hline
\end{tabular}
\caption{Several spectral weights in units of $\mu_B^2$ as defined in the text for several $U/t$ values and
some results from Refs. \cite{lorenzana2005,Dionys-09}.}
\label{table2}
\end{table} 

From our above analysis, the amount of spin spectral weight for excitation energy 
$\hbar\omega >$ 450 meV is small, $W_{>450}\approx 0.1\,\mu_B^2$. By combining
our spectral-weight results with the boundaries of the spin-triplet spectrum
plotted in Fig. \ref{fig4}, such a small spin spectral weight is expected to extend to about 566 meV,
mostly at and around the momentum $[\pi,\pi]$.

\section{Concluding remarks}
\label{concluding}

In this paper we have studied by means of the half-filled Hubbard model on the square lattice
several open issues raised by the recent LCO neutron scattering data reported in Ref. \cite{headings2010}. 
Our studies combined standard methods such as RPA techniques involving a broken symmetry
ground state with DMRG calculations on Hubbard cylinders and a spinon representation
suitable to the LCO intermediate interaction range $U/t\in (6,8)$. 
The latter emerges from a rotated-electron operator representation that
has been constructed to ensure that rotated-electron
single and double occupancy are good quantum numbers for $U/t>0$.
This assures that our spinons are well defined for the LCO intermediate $U/t$ range. 
Indeed these spinons are the spins of the rotated electrons that singly
occupy sites. In the large-$U/t$ limit the rotated-electrons become electrons, so that
one recovers the usual picture.

Within this operator formulation, the Hubbard model in the VDU subspace considered
here can be mapped onto a spin-only problem for the $U/t$ range of interest
for our studies. The spin excitations
preserve the electron number. At fixed electron number the VDU subspace is the 
only one within a finite excitation-energy 
window, $\omega\in (0,2\Delta_{MH})$. Here $2\Delta_{MH}$ is the Mott-Hubbard gap,
whose $U/t$ dependence we have calculated by DMRG. We have found that
$2\Delta_{MH}\in (816$ meV$,1442$ meV) for $U/t\in (6,8)$. 
Correspondingly, the largest energy of the general spin spectrum of the half-filled
Hubbard model on the square lattice in the VDU subspace
represented in Fig. \ref{fig4} is $566$ meV, so that it is 
fully contained in this energy window.

The coherent part of the spin spectrum, which corresponds to the spin-wave spectrum,
was complementarily studied by a RPA analysis involving a broken symmetry
ground state and the spinon operator representation. The former method was also used
to calculate the spin-wave intensity momentum distribution. Quantitative agreement with both the spin-wave spectrum
obtained by summing up an infinite number of ladder diagrams and that observed
in LCO neutron scattering experimental studies is reached by the spinon method
provided that the initial $n=1$ and $m=0$ ground state has $d$-wave spinon pairing.
For that ground state such a pairing refers only to the rotated-electron spin degrees
of freedom. Whether upon hole doping such a pre-formed $d$-wave spinon pairing could
lead to rotated-electron $d$-wave pairing or even related electron $d$-wave pairing 
is an issue that deserves further investigations.

Following the good quantitative agreement with the spin-wave spectrum, the spinon
representation was used to derive the full $S_s=1$ spin-triplet spectrum
represented in Fig. \ref{fig4} for the $U$ and $t$ values suitable for LCO.
From analysis of that figure we have found that for the momentum $[\pi,0]$ there are no excited states of energy higher than
the spin waves. Thus at momentum $[\pi,0]$, the continuum weight distribution energy-integrated 
intensity vanishes or is extremely small. Such an intensity is expected to follow the
same trend as that of the spin waves. Hence this behavior is consistent with a corresponding damping of the spin-wave 
intensity at $[\pi,0]$ observed in the recent high-energy inelastic
neutron-scattering experiments of Ref. \cite{headings2010}. 

On the other hand, a resonant-inelastic x-ray scattering study of insulating and doped La$_{2-x}$Sr$_x$CuO$_4$
found a mode at $500$ meV, at a momentum transfer $[\pi,0]$ \cite{LSCO-pi-0}.
This $500$ meV mode is observed only when the incident x-ray polarization is normal to the CuO planes.
It could be a d-d crystal-field excitation \cite{dd-93,dd-94}, rather than a spin excitation. In case 
it is a spin excitation, one possible explanation given in Ref. \cite{LSCO-pi-0} is that it involves two 
spin-flip processes, created on adjacent copper-oxide planes. Since our present study relies on the 
Hubbard model on a single square-lattice plane, that mechanism would be beyond our theoretical approach.

We recall that for each excitation momentum $\vec{k}$, states in the $(\vec{k},\omega)$ domain of Fig. \ref{fig4} 
whose energy is lower than that of the spin-wave spectrum intermediate sheet generate no
spectral weight and thus do not contribute 
to the spin dynamical structure factor. Furthermore, consistent with the first-moment sum rules 
of an isotropic antiferromagnet, no and nearly no weight is generated by states 
of any energy and momentum $[0,0]=[0,2\pi]=[2\pi,0]=[2\pi,2\pi]$ and near it, respectively. That together
with the small amount of spin spectral weight reported in Table \ref{table2} for energies between $450$ meV and $566$ meV
indicates that in that energy window there is nearly
no spin spectral weight near the momentum $[0,0]=[0,2\pi]=[2\pi,0]=[2\pi,2\pi]$ (see Fig. \ref{fig4}).

In addition to the Mott-Hubbard gap magnitude dependence on $U/t$, DMRG calculations were performed to derive 
the $U/t$ dependence of the ground-state electron single occupancy expectation value $(1-d)$. That quantity
plays an important role in our study, in that it controls several spin-weight sum rules. 
Our prediction for the amount of total spin spectral weight in the energy range 
$\omega\in (0$ meV$,450$ meV) quantitatively agrees with that observed in the recent
high-energy inelastic neutron scattering studies of Ref. \cite{headings2010}, which were
limited to that energy window. 

Moreover, as reported in Table \ref{table2} we predict that there is a small amount of extra weight $\approx 0.1\,\mu_B^2$ 
above $450$ meV, which extends to about $566$ meV. Since at and near the momentum $[0,0]=[0,2\pi]=[2\pi,0]=[2\pi,2\pi]$ there 
is nearly no spin spectral weight, analysis of Fig. \ref{fig4} reveals that for energies between $450$ meV and $566$ meV the small amount
of extra spin spectral weight is located at and around the momentum $[\pi,\pi]$.
Thus we suggest that future LCO neutron scattering experiments scan the energies between $450$ meV and 
$566$ meV and momenta around $[\pi,\pi]$.

\acknowledgements
We thank the authors of Ref. \cite{headings2010} for
providing their experimental data and A. Muramatsu for discussions. 
J. M. P. C. thanks the hospitality of the University of Stuttgart
and the support of the Portuguese FCT under SFRH/BSAB/1177/2011, German Transregional Collaborative 
Research Center SFB/TRR21, and Max Planck Institute for Solid State Research. 
S.R.W. acknowledges the support of the NSF under DMR 090-7500.

\appendix

\section{Useful operators algebra}
\label{commute}

Here we justify why the six generators of the global $\eta$-spin and spin $SU(2)$ symmetries commute 
with the electron - rotated-electron unitary operator. Moreover, we address the problem of
the $c$ fermion operator, spinon operator, and $\eta$-spinon operator algebras.

To achieve our first goal, it is useful to express the kinetic-energy operator $\hat{T}$ given in Eq. (\ref{HH}) 
as $\hat{T}= \hat{T}_0 + \hat{T}_{+1} + \hat{T}_{-1}$. Here,
\begin{eqnarray}
\hat{T}_{\gamma} & = & -\sum_{\langle j,j'\rangle}\hat{T}_{\gamma;,j,j'}  \, ,
\hspace{0.25cm} \gamma = 0, \pm 1 \, ,
\nonumber \\
\hat{T}_{0;j,j'} & = & \sum_{\sigma}[\hat{n}_{\vec{r}_j,-\sigma}\,c_{\vec{r}_j,\sigma}^{\dag}\,
c_{\vec{r}_{j'},\sigma}\,\hat{n}_{\vec{r}_{j'},-\sigma} 
\nonumber \\
& + & (1-\hat{n}_{\vec{r}_j,-\sigma})\,c_{\vec{r}_j,\sigma}^{\dag}\,
c_{\vec{r}_{j'},\sigma}\,(1-\hat{n}_{\vec{r}_{j'},-\sigma}) +c.c.] \, ,
\nonumber \\
\hat{T}_{+1;j,j'} & = & 
\sum_{\sigma}[\hat{n}_{\vec{r}_j,-\sigma}\,c_{\vec{r}_j,\sigma}^{\dag}\,c_{\vec{r}_{j'},\sigma}\,(1-\hat{n}_{\vec{r}_{j'},-\sigma}) 
\nonumber \\
& + & \hat{n}_{\vec{r}_{j'},-\sigma}\,c_{\vec{r}_{j'},\sigma}^{\dag}\,c_{\vec{r}_{j},\sigma}\,(1-\hat{n}_{\vec{r}_{j},-\sigma})] \, ,
\nonumber \\
\hat{T}_{-1;j,j'} & = & \sum_{\sigma}[(1-\hat{n}_{\vec{r}_j,-\sigma})\,c_{\vec{r}_j,\sigma}^{\dag}\,
c_{\vec{r}_{j'},\sigma}\,\hat{n}_{\vec{r}_{j'},-\sigma} 
\nonumber \\
& + & (1-\hat{n}_{\vec{r}_{j'},-\sigma})\,c_{\vec{r}_{j'},\sigma}^{\dag}\,
c_{\vec{r}_{j},\sigma}\,\hat{n}_{\vec{r}_{j},-\sigma}] \, .
\label{T-op}
\end{eqnarray}
While the operator $\hat{T}_0$ does not change electron double occupancy, the operators $\hat{T}_{+1}$ and $\hat{T}_{-1}$ 
change it by $+1$ and $-1$, respectively. 

For $U/t>0$ the operator $\hat{S}$ in the expression
${\hat{V}} = e^{-{\hat{S}}}$ given in Eq. (\ref{rotated-operators}) can be expanded in a series of $t/U$,
\begin{equation}
{\hat{S}} = -{t\over U}\,\left[\hat{T}_{+1} -\hat{T}_{-1}\right] 
+ {\cal{O}} (t^2/U^2) \, .
\label{OOr}
\end{equation} 
Although as discussed in Ref. \cite{bipartite} there are infinite choices for the operators ${\hat{V}} = e^{-{\hat{S}}}$ and $\hat{S}$,
they share two important properties \cite{bipartite,Stein,HO-04}: (i) To leading order in $t/U$ all 
read $-{t\over U}\,[\hat{T}_{+1} -\hat{T}_{-1}]$, as given in Eq. (\ref{OOr}); (ii) 
Their operational expressions involve only the kinetic operators $\hat{T}_0$, $\hat{T}_{+1}$, and $\hat{T}_{-1}$
of Eq. (\ref{T-op}). Such properties apply to the specific electron - rotated-electron unitary operator ${\hat{V}}$
uniquely defined in this paper. 

The rotated kinetic operators $\tilde{T}_0$, $\tilde{T}_{+1}$, and $\tilde{T}_{-1}$ such that
$\tilde{T}_{\gamma}={\hat{V}}^{\dag}\,\hat{T}_{\gamma}\,{\hat{V}}$ for $\gamma =0,\pm 1$ are given by,
\begin{eqnarray}
\tilde{T}_{\gamma} & = & -\sum_{\langle j,j'\rangle}\tilde{T}_{\gamma;,j,j'}  \, ,
\hspace{0.25cm} \gamma = 0, \pm 1 \, ,
\nonumber \\
\tilde{T}_{0;j,j'} & = & \sum_{\sigma}[\tilde{n}_{\vec{r}_j,-\sigma}\,\tilde{c}_{\vec{r}_j,\sigma}^{\dag}\,
\tilde{c}_{\vec{r}_{j'},\sigma}\,\tilde{n}_{\vec{r}_{j'},-\sigma} 
\nonumber \\
& + & (1-\tilde{n}_{\vec{r}_j,-\sigma})\,\tilde{c}_{\vec{r}_j,\sigma}^{\dag}\,
\tilde{c}_{\vec{r}_{j'},\sigma}\,(1-\tilde{n}_{\vec{r}_{j'},-\sigma})+c.c.] \, ,
\nonumber \\
\tilde{T}_{+1;j,j'} & = & 
\sum_{\sigma}[\tilde{n}_{\vec{r}_j,-\sigma}\,\tilde{c}_{\vec{r}_j,\sigma}^{\dag}\,\tilde{c}_{\vec{r}_{j'},\sigma}\,(1-\tilde{n}_{\vec{r}_{j'},-\sigma}) 
\nonumber \\
& + & \tilde{n}_{\vec{r}_{j'},-\sigma}\,\tilde{c}_{\vec{r}_{j'},\sigma}^{\dag}\,\tilde{c}_{\vec{r}_{j},\sigma}\,(1-\tilde{n}_{\vec{r}_{j},-\sigma})] \, ,
\nonumber \\
\tilde{T}_{-1;j,j'} & = & 
\sum_{\sigma}[(1-\tilde{n}_{\vec{r}_j,-\sigma})\,\tilde{c}_{\vec{r}_j,\sigma}^{\dag}\,
\tilde{c}_{\vec{r}_{j'},\sigma}\,\tilde{n}_{\vec{r}_{j'},-\sigma} 
\nonumber \\
& + & (1-\tilde{n}_{\vec{r}_{j'},-\sigma})\,\tilde{c}_{\vec{r}_{j'},\sigma}^{\dag}\,
\tilde{c}_{\vec{r}_{j},\sigma}\,\tilde{n}_{\vec{r}_{j},-\sigma}] \, .
\label{T-op-rot}
\end{eqnarray}

To confirm that the three generators of the spin $SU(2)$ symmetry, three generators
of the $\eta$-spin $SU(2)$ symmetry, and also the momentum operator $\hat{P}$
commute with the electron - rotated-electron unitary operator ${\hat{V}}={\tilde{V}}$, 
one uses the exact result that the unitary operator ${\hat{V}}$ can be
solely expressed in terms of the three kinetic operators given in Eq. (\ref{T-op}) \cite{bipartite,Stein}. 
In Ref. \cite{bipartite} the following 21 commutators were found to vanish,  
\begin{eqnarray}
[\hat{P},\hat{T}_{\gamma}] & = &
[{\hat{S}}_{\alpha}^z,\hat{T}_{\gamma}] = [{\hat{S }}_{\alpha}^{\dagger},\hat{T}_{\gamma}] = [{\hat{S }}_{\alpha},\hat{T}_{\gamma}] =0 \, ,
\nonumber \\
\alpha & = & \eta , s \, , \hspace{0.15cm} \gamma =0,\pm 1 \, .
\label{S-T}
\end{eqnarray}
Although the algebra involved in their derivation is cumbersome, it is straightforward. The vanishing of the 
commutators given in Eq. (\ref{S-T}) then implies that the momentum operator and the six generators of the $\eta$-spin 
and spin algebras commute with the unitary operator ${\hat{V}}$,
\begin{eqnarray}
[\hat{P},{\hat{V}}] & = &
[{\hat{S}}_{\alpha}^z,{\hat{V}}] = [{\hat{S }}_{\alpha}^{\dagger},{\hat{V}}] = [{\hat{S }}_{\alpha},{\hat{V}}] =0 \, ,
\nonumber \\
\alpha & = & \eta , s \, , \hspace{0.15cm} l=0,\pm 1 \, .
\label{S-V-dag}
\end{eqnarray}

Hence the above operators have
the same expression in terms of electron and rotated-electron 
creation and annihilation operators, so that the momentum operator reads,
\begin{equation}
\hat{{\vec{P}}}  = \sum_{\sigma=\uparrow ,\downarrow }\sum_{\vec{k}}\,\vec{k}\,
c_{\vec{k},\sigma }^{\dag }\,c_{\vec{k},\sigma } =
\sum_{\sigma=\uparrow ,\downarrow }\sum_{\vec{k}}\,\vec{k}\,
{\tilde{c}}_{\vec{k},\sigma }^{\dag }\,{\tilde{c}}_{\vec{k},\sigma } \, .
\label{P-invariant}
\end{equation}
Furthermore, the above-mentioned six generators are given by,
\begin{eqnarray}
{\hat{S}}_{\eta}^{z} & = & \sum_{j=1}^{N_a}{\hat{s}}_{\vec{r}_j,\eta}^{z}
= \sum_{j=1}^{N_a}{\tilde{s}}_{\vec{r}_j,\eta}^{z} \, ,
\nonumber \\
{\hat{S}}_{\eta}^{\dag} & = & \sum_{j=1}^{N_a}{\hat{s}}_{\vec{r}_j,\eta}^{+}
= \sum_{j=1}^{N_a}{\tilde{s}}_{\vec{r}_j,\eta}^{+} \, ,
\nonumber \\
{\hat{S}}_{\eta} & = & \sum_{j=1}^{N_a}{\hat{s}}_{\vec{r}_j,\eta}^{-}
= \sum_{j=1}^{N_a}{\tilde{s}}_{\vec{r}_j,\eta}^{-} \, ,
\nonumber \\
{\hat{S}}_{s}^{z} & = & \sum_{j=1}^{N_a}{\hat{s}}_{\vec{r}_j,s}^{z}
= \sum_{j=1}^{N_a}{\tilde{s}}_{\vec{r}_j,s}^{z} \, ,
\nonumber \\
{\hat{S}}_{s}^{\dag} & = & \sum_{j=1}^{N_a}{\hat{s}}_{\vec{r}_j,s}^{+}
= \sum_{j=1}^{N_a}{\tilde{s}}_{\vec{r}_j,s}^{+} \, ,
\nonumber \\
{\hat{S}}_{s} & = & \sum_{j=1}^{N_a}{\hat{s}}_{\vec{r}_j,s}^{-}
= \sum_{j=1}^{N_a}{\tilde{s}}_{\vec{r}_j,s}^{-} \, .
\label{Scs}
\end{eqnarray}
Those of the unrotated local operators appearing here associated with the
$\eta$-spin algebra read,
\begin{eqnarray}
{\hat{s}}_{\vec{r}_j,\eta}^{z} & = & -{1\over 2}[1-{\hat{n}}_{\vec{r}_j,\uparrow}-{\hat{n}}_{\vec{r}_j,\downarrow}]  \, ,
\nonumber \\
{\hat{s}}_{\vec{r}_j,\eta}^{+} & = & e^{i\vec{\pi}\cdot\vec{r}_j}\,c_{\vec{r}_j,\downarrow}^{\dag}\,
c_{\vec{r}_j,\uparrow}^{\dag}  \, ,
\nonumber \\
{\hat{s}}_{\vec{r}_j,\eta}^{-} & = & e^{-i\vec{\pi}\cdot\vec{r}_j}\,c_{\vec{r}_j,\uparrow}\,c_{\vec{r}_j,\downarrow} 
\, , \hspace{0.25cm} j = 1,2,...,N_a \, ,
\label{S-j-eta}
\end{eqnarray}
whereas those associated with the spin algebra are given in Eq. (\ref{S-j-s}).
On the other hand, the rotated local operators appearing in the alternative expressions of
Eq. (\ref{Scs}) are provided in Eq. (\ref{Scs-j-rot}).

The six local operators given in Eqs. (\ref{S-j-s}) and (\ref{S-j-eta}) together with the local operator,
\begin{equation}
{\hat{s}}_{\vec{r}_j,c} = \sum_{\sigma =\uparrow
,\downarrow}\,{\hat{n}}_{\vec{r}_j,\sigma}\,(1- {\hat{n}}_{\vec{r}_j,-\sigma}) 
\, , \hspace{0.25cm} j = 1,2,...,N_a \, ,
\label{S-j-c}
\end{equation}
are the seven generators of the $U\neq 0$ local gauge $SU(2)\times SU(2) \times U(1)$ symmetry 
of the Hubbard model on a bipartite lattice with vanishing transfer integral, $t=0$ \cite{U(1)-NL}.

Since the electron - rotated-electron transformation generated by the operator $\hat{V}$
is unitary, the rotated-electron operators ${\tilde{c}}_{\vec{r}_j,\sigma}^{\dag}$ 
and ${\tilde{c}}_{\vec{r}_j,\sigma}$ of Eq. (\ref{rotated-operators})
have the same anticommutation relations as the corresponding electron 
operators $c_{\vec{r}_j,\sigma}^{\dag}$ and $c_{\vec{r}_j,\sigma}$, respectively.
Similarly, the local $c$ fermion operators of Eq. (\ref{fc+}) and three local spinon operators and three
local $\eta$-spinon operators of Eq. (\ref{Scs-j-rot}) have the same algebra as the
corresponding unrotated spin-less and $\eta$-spin-less fermion operators of Eq. (\ref{fc-unrot})
and and three local spin operators of Eq. (\ref{S-j-s}) and three
local $\eta$-spin operators of Eq. (\ref{S-j-eta}), respectively. The former
operators play a major role in the finite-$U/t$ physics of the model. The latter
operators are a limiting case of the former operators reached for
$U/t\gg 1$. Hence, without loss of generality in the following we provide the algebra
of the local $c$ fermion operators of Eq. (\ref{fc+}) and three $\eta s$
quasi-spin operators of Eq. (\ref{rotated-quasi-spin}). The $SU(2)$ algebra of the
latter three operators fully determines those of the three local spinon operators and three
local $\eta$-spinon operators of Eq. (\ref{Scs-j-rot}).

Straightforward manipulations based on Eqs. (\ref{fc+})-(\ref{rotated-quasi-spin}) lead
to the following algebra for the $c$ fermion operators,
\begin{eqnarray}
\{f^{\dag}_{\vec{r}_j,c}\, ,f_{\vec{r}_{j'},c}\} & = & \delta_{j,j'}  \, ,
\nonumber \\
\{f_{\vec{r}_j,c}^{\dag}\, ,f_{\vec{r}_{j'},c}^{\dag}\} & = &
\{f_{\vec{r}_j,c}\, ,f_{\vec{r}_{j'},c}\} = 0 \, ,
\label{albegra-cf}
\end{eqnarray}
and the $c$ fermion operators and the local $\eta s$ quasi-spin operators,
\begin{eqnarray}
\left[f_{\vec{r}_j,c}^{\dag}\, ,{\tilde{q}}^l_{\vec{r}_{j'}}\right] & = &
\left[f_{\vec{r}_j,c}\, ,{\tilde{q}}^l_{\vec{r}_{j'}}\right] = 0 \, ,
\nonumber \\
\left[f_{\vec{r}_j,c}^{\dag}\, ,{\tilde{s}}^l_{\vec{r}_{j'},\alpha}\right] & = &
\left[f_{\vec{r}_j,c}\, ,{\tilde{s}}^l_{\vec{r}_{j'},\alpha}\right] = 0 \, ,
\nonumber \\ 
l & = & \pm,x_3 \, , \hspace{0.25cm} \alpha =\eta , s \, .
\label{albegra-cf-s-h}
\end{eqnarray}

The $SU(2)$ algebra obeyed by the local $\eta s$ quasi-spin operators ${\tilde{q}}^{l}_{\vec{r}_j}$ 
where $l = x_3, \pm$, such that 
${\tilde{q}}^{\pm}_{\vec{r}_j}= {\tilde{q}}^{x}_{\vec{r}_j}\pm i\,{\tilde{q}}^{y}_{\vec{r}_j}$, 
and corresponding $\eta$-spinon ($\alpha =\eta$) and spinon ($\alpha =s$) operators 
${\tilde{s}}^{l}_{\vec{r}_j,\alpha}$ is,
\begin{equation}
\left[{\tilde{q}}^{+}_{\vec{r}_j},{\tilde{q}}^{-}_{\vec{r}_{j'}}\right] = \delta_{j,j'}\,2\,{\tilde{q}}^{x_3}_{\vec{r}_j}
\, ; \hspace{0.35cm}
\left[{\tilde{q}}^{\pm}_{\vec{r}_j},{\tilde{q}}^{x_3}_{\vec{r}_{j'}}\right] = \mp \delta_{j,j'}\,{\tilde{q}}^{\pm}_{\vec{r}_j} \, ,
\label{albegra-q-com}
\end{equation}
and
\begin{eqnarray}
\left[{\tilde{s}}^{+}_{\vec{r}_j,\alpha},{\tilde{s}}^{-}_{\vec{r}_{j'},\alpha'}\right] & = & \delta_{j,j'}\delta_{\alpha,\alpha'}\,2\,{\tilde{s}}^{x_3}_{\vec{r}_j,\alpha} \, ,
\nonumber \\
\left[{\tilde{s}}^{\pm}_{\vec{r}_j,\alpha},{\tilde{s}}^{x_3}_{\vec{r}_{j'},\alpha'}\right] & = & \mp \delta_{j,j'}\delta_{\alpha,\alpha'}\,{\tilde{s}}^{\pm}_{\vec{r}_j,\alpha} \, ,
\nonumber \\
\alpha ,\alpha' & = & \eta ,s \, ,
\label{albegra-s-e-s-com}
\end{eqnarray}
respectively. Moreover, one has obviously that $[{\tilde{q}}^{l}_{\vec{r}_j},{\tilde{q}}^{l}_{\vec{r}_{j'}}]=0$
and $[{\tilde{s}}^{l}_{\vec{r}_j,\alpha},{\tilde{s}}^{l}_{\vec{r}_{j'},\alpha'}] = 0$ where $l= 0, \pm$ and
$\alpha ,\alpha'=\eta , s$. While the $c$ fermion and $\eta s$ quasi-spin operator algebras refer to the whole Hilbert space, those of the
$\eta$-spinon and spinon operators correspond to well-defined subspaces spanned by states whose 
value of the number $2S_c$ of rotated-electron singly occupied sites is fixed. This ensures
that the value of the corresponding $\eta$-spinon number $M_{\eta}=[N_a -2S_c]$ and spinon number $M_{s}=2S_c$ 
is fixed as well. 

The relations given in Eqs. (\ref{albegra-cf})-(\ref{albegra-s-e-s-com}) confirm that when acting onto 
the model's Hilbert space the $c$ fermions associated 
with the global $c$ hidden $U(1)$ symmetry are $\eta$-spinless and spinless fermionic objects. They are consistent 
as well with the spinons and $\eta$-spinons being spin-$1/2$ and $\eta$-spin-$1/2$ objects, respectively, 
whose local operators obey the usual corresponding $SU(2)$ algebras. 

\section{Spin-wave spectrum in the high symmetry directions}
\label{SS-spectrum}

In this appendix we study the spin-wave spectrum of Eq. (\ref{om-SW})
in the BZ high symmetry directions. These directions 
correspond to those measured by high-resolution inelastic 
neutron scattering in LCO, as plotted for instance in Fig. 3 (A) of Ref. \cite{LCO-2001}.
(Our theoretical spin-wave spectrum curves are plotted along with the more recent
LCO high-energy neutron scattering points of Ref. \cite{headings2010} in Fig. \ref{fig2}.)

We denote such symmetry directions by $MO$, $\Gamma O$, $XM$, $\Gamma X$, 
and $XO$. They connect the momentum-space points $M=[\pi,\pi]$, $O=[\pi/2,\pi/2]$, 
$\Gamma =[0,0]$, and $X=[\pi,0]$ of the general spin-wave spectrum 
provided in Eq. (\ref{om-SW}). The use of that equation reveals that the spin-wave 
excitation spectrum is in such symmetry directions given by, 
\begin{eqnarray}
\omega_{\Gamma O} (\vec{k}) & = & {\mu^0\over 2}\sin (k_i) \, ,
\nonumber \\
\vec{k} & = & [\pi,-\pi] - \vec{q}  - \vec{q}\,'
\nonumber \\
& = & [k_i,k_i] \, , \hspace{0.25cm} k_i = k_x = k_y \in (0,\pi/2) \, ,
\label{direc-GO}
\end{eqnarray}
for $s1$ fermion hole momenta,
\begin{eqnarray}
\vec{q} & = & [\pi/2 -k_i,-\pi/2 -k_i] \, , \hspace{0.25cm} k_i \in (0,\pi/2) \, ,
\nonumber \\
\vec{q}\,' & = & [\pi/2,-\pi/2] \, ,
\label{qq-direc-GO}
\end{eqnarray}
\begin{eqnarray}
\omega_{MO} (\vec{k}) & = & {\mu^0\over 2}\sin (k_i) \, ,
\nonumber \\
\vec{k} & = & [\pi,\pi] - \vec{q} - \vec{q}\,'
\nonumber \\
& = & [k_i,k_i] \, , \hspace{0.25cm} k_i = k_x = k_y \in (\pi/2,\pi) \, ,
\label{direc-MO}
\end{eqnarray}
for $s1$ fermion hole momenta,
\begin{eqnarray}
\vec{q} & = & [\pi/2 -k_i,3\pi/2 -k_i] \, , \hspace{0.25cm} k_i \in (\pi/2,\pi) \, ,
\nonumber \\
\vec{q}\,' & = & [\pi/2,-\pi/2] \, ,
\label{qq-direc-MO}
\end{eqnarray}
\begin{eqnarray}
\omega_{\Gamma X} (\vec{k}) & = &
\left[{\mu^0\over 2} +W^0_{s1}\right]\sin (k_x/2) \, ,
\nonumber \\
\vec{k} & = & [\pi,-\pi] - \vec{q} - \vec{q}\,'
\nonumber \\
& = & [k_x,0]   \, , \hspace{0.25cm} k_x \in (0,\pi) \, ,
\label{direc-GX}
\end{eqnarray}
for $s1$ fermion hole momenta,
\begin{eqnarray}
\vec{q} & = & [\pi/2-k_x/2,-\pi/2 -k_x/2] \, , \hspace{0.25cm} k_x \in (0,\pi) \, ,
\nonumber \\
\vec{q}\,' & = & [\pi/2 -k_x/2,-\pi/2+k_x/2] \, , \hspace{0.25cm} k_x \in (0,\pi) \, ,
\label{qq-direc-GX}
\end{eqnarray}
\begin{eqnarray}
\omega_{XM} (\vec{k}) & = &
\left[{\mu^0\over 2} +W^0_{s1}\right]\cos (k_y/2) \, ,
\nonumber \\
\vec{k} & = & [\pi,\pi] - \vec{q} - \vec{q}\,' 
\nonumber \\
& = & [\pi,k_y]  \, , \hspace{0.25cm} k_y \in (0,\pi) \, ,
\label{direc-XM}
\end{eqnarray}
for $s1$ fermion hole momenta,
\begin{eqnarray}
\vec{q} & = &  [-k_y/2,\pi -k_y/2] \, , \hspace{0.25cm} k_y \in (0,\pi) \, ,
\nonumber \\
\vec{q}\,' & = & [k_y/2,-k_y/2] \, , \hspace{0.25cm} k_y \in (0,\pi) \, ,
\label{qq-direc-XM}
\end{eqnarray}
and,
\begin{eqnarray}
\omega_{XO} (\vec{k}) & = &
{\mu^0\over 2} -W^0_{s1} \cos (k_x) \nonumber \\
& = & {\mu^0\over 2} +W^0_{s1} \cos (k_y) \, ,
\nonumber \\
\vec{k}  & = & [\pi,-\pi] - \vec{q} - \vec{q}\,'
\nonumber \\
& = & [\pi,\pi] - \vec{q}\,'' - \vec{q}\,'''
\nonumber \\
& = &  [k_x,\pi -k_x] 
\, , \hspace{0.25cm} k_x \in (\pi/2,\pi)
\nonumber \\
& = &  [\pi -k_y,k_y] 
\, , \hspace{0.25cm} k_y \in (0,\pi/2) \, ,
\label{direc-XO}
\end{eqnarray}
for $s1$ fermion hole momenta,
\begin{eqnarray}
\vec{q} & = &  [0,-\pi] \, ,
\nonumber \\
\vec{q}\,' & = & [\pi -k_x,-\pi +k_x] \, , \hspace{0.25cm} k_x \in (\pi/2,\pi) \, ,
\label{qq-direc-XO-A}
\end{eqnarray}
or,
\begin{eqnarray}
\vec{q}\,'' & = & [0,\pi]  \, ,
\nonumber \\
\vec{q}\,''' & = & [k_y,-k_y] \, , \hspace{0.25cm} k_y \in (0,\pi/2) \, ,
\label{qq-direc-XO-B}
\end{eqnarray}
respectively.

The theoretical spin excitation spectra, Eqs. (\ref{direc-GO})-(\ref{direc-XO}),
are plotted in Fig. \ref{fig5} (solid line) for $U/t\approx 6.1$
and $t\approx 0.295$ eV together with the experimental results of Ref. \cite{LCO-2001} (circles) 
for $T=10$ K. Such $U/t$ and $t$ magnitudes correspond to 
$\mu^0= 565.6$ meV and $W^0_{s1}= 49.6$ meV in the above
energy spectra. The spin-spectrum expressions provided in Eqs. (\ref{direc-GO})-(\ref{direc-XO})  
refer to the first BZ. In Fig. \ref{fig5} they are plotted in the second
BZ, alike in Fig. 3 (A) of Ref. \cite{LCO-2001}.
An excellent quantitative agreement is reached for the above magnitudes of the involved
energy scales.


\begin{thebibliography}{99}

\bibitem{WSstripe}
See, for example,
S. R. White and D. J. Scalapino, 
Phys. Rev. B {\bf 79}, 220504 (R) (2009).

\bibitem{headings2010}
N. S. Headings, S. M. Hayden, R. Coldea, and T. G. Perring, 
Phys. Rev. Lett. {\bf 105}, 247001 (2010).

\bibitem{SWZ}
J. R. Schrieffer, X.G. Wen, and S. C. Zhang, 
Phys. Rev. B {\bf 39}, 11663 (1989).

\bibitem{LCO-2001} 
R. Coldea, S. M. Hayden, G. Aeppli, T. G. Perring, C. D. Frost, T. E. Mason, 
S.-W. Cheong, and Z. Fisk, Phys. Rev. Lett. {\bf 86}, 5377 (2001).

\bibitem{peres2002} 
N. M. R. Peres and  M. A. N. Ara\'ujo,
Phys. Rev. B {\bf 65}, 132404 (2002).

\bibitem{lorenzana2005}
J. Lorenzana, G. Seibold, and R. Coldea,
Phys. Rev. B {\bf 72}, 224511 (2005).

\bibitem{companion}
J. M. P. Carmelo, Nucl. Phys. B {\bf 824}, 452 (2010);
J. M. P. Carmelo, Nucl. Phys. B {\bf 840}, 553 (2010), Erratum.

\bibitem{Tremblay-09} 
J.-Y. P. Delannoy, M. J. P. Gingras, P. C. W. Holdsworth, and A.-M. S. Tremblay,
Phys. Rev. B {\bf 79}, 235130 (2009).

\bibitem{Miguel-03} 
N. M . Peres and  M. A. N. Ara\'ujo,
Physica Stat. Sol. {\bf 236}, 523 (2003).

\bibitem{DMRG-SL-2011}
S. Yan, D. A. Huse, and S. R. White,
Science {\bf 332}, 1173 (2011).

\bibitem{DMRG-tJ-2011}
P. Corboz, S. R. White, G. Vidal, and M. Troyer,
Phys. Rev. B {\bf 84}, 041108 (2011).

\bibitem{Steve-92}
S. R. White, Phys. Rev. Lett. {\bf 69}, 2863 (1992).

\bibitem{companion0}
J. M. P. Carmelo, Ann. Phys. {\bf 327},  553 (2012). 

\bibitem{bipartite}
J. M. P. Carmelo, S. \"Ostlund, and M. J. Sampaio,
Ann. Phys. {\bf 325}, 1550 (2010).

\bibitem{HL}
O. J. Heilmann and E. H. Lieb, Ann. N. Y. Acad. Sci. {\bf 172}, 583 (1971).

\bibitem{Lieb89}
E. H. Lieb, Phys. Rev. Lett. {\bf 62}, 1201 (1989).

\bibitem{Yang89}
C. N. Yang, Phys. Rev. Lett. {\bf 63}, 2144 (1989).

\bibitem{Zhang}
C. N. Yang and S. C. Zhang, Mod. Phys. Lett. B {\bf 4} 759 (1990);    
S. C. Zhang, Phys. Rev. Lett. {\bf 65}, 120 (1990).

\bibitem{U(1)-NL} 
S. \"Ostlund, E. Mele, Phys. Rev. B {\bf 44}, 12413 (1991).

\bibitem{Mano} 
E. Manousakis,
Rev. Mod. Phys. {\bf 63}, 1 (1991).

\bibitem{Dionys-09}
D. Baeriswyl, D. Eichenberger, and M. Menteshashvili,
New J. Phys. {\bf 11}, 075010 (2009).

\bibitem{BR-70} 
W. F. Brinkman and T. M. Rice,
Phys. Rev. B {\bf 2}, 4302 (1970).	

\bibitem{MZ-89}
Walter Metzner and Dieter Vollhardt,
Phys. Rev. B {\bf 39}, 4462 (1989).

\bibitem{Paiva}
For a finite temperature quantum Monte Carlo estimation of $1-2d$, see
T. Paiva, R.T. Scalettar, C. Huscroft, and A.K. McMahan, 
Phys. Rev. B {\bf 63}, 125116 (2001).

\bibitem{brinckmann-lee} J. Brinckmann and P. A. Lee,
 Phys. Rev. Lett. {\bf 82}, 2915 (1999).

\bibitem{Canali}
C. M. Canali, S. M. Girvin, and Mats Wallin,
Phys. Rev. B {\bf 45}, 10 131 (1992);
C. M. Canali and Mats Wallin,
Phys. Rev. B {\bf 48}, 3264 (1993).

\bibitem{authors}
S. M. Hayden and R. Coldea, private communication.

\bibitem{Ageluci}
A. Angelucci, Phys. Rev. B {\bf 51}, 11580 (1995).	

\bibitem{Mura-03}
C. Lavalle, M. Arikawa, S. Capponi, F. F. Assaad, and A. Muramatsu, 
Phys. Rev. Lett. {\bf 90}, 216401 (2003). 

\bibitem{Stellan-06}
S. \"Ostlund and M. Granath, Phys. Rev. Lett {\bf 96}, 066404 (2006).

\bibitem{Stein} 
J. Stein, J. Stat. Phys. {\bf 88}, 487 (1997).

\bibitem{HO-04}
A. L. Chernyshev, D. Galanakis, P. Phillips, A. V. Rozhkov, and A.-M. S. Tremblay,
Phys. Rev. B {\bf 70}, 235111 (2004).	

\bibitem{Mac}
A. H. MacDonald, S. M. Girvin, and D. Yoshioka,
Phys. Rev. B {\bf 41}, 2565 (1990); {\bf 37}, 9753 (1988) .

\bibitem{Harris}
A. B. Harris and R. V. Lange,
Phys. Rev. {\bf 157}, 295 (1967).	

\bibitem{Moskvin-11} 
A. S. Moskvin, Phys. Rev. B {\bf 84}, 075116 (2011).
	
\bibitem{Ono-07} 
S. Ono, S. Komiya, and Y. Ando,  
Phys. Rev. B {\bf 75}, 024515 (2007).

\bibitem{Peter-88}
P. Horsch and W. von der Linden,
Z. Phys. B {\bf 72}, 181 (1988).      

\bibitem{1D-03}
J. M. P. Carmelo and P. D. Sacramento, 
Phys. Rev. B {\bf 68}, 085104 (2003).

\bibitem{1D-04}
J. M. P. Carmelo, J. M. Rom\'an, and K. Penc, 
Nucl. Phys. B {\bf 683}, 387 (2004).

\bibitem{old-88}
S. Liang, B. Dou\c{c}ot, and P. W. Anderson, 
Phys. Rev. Lett. {\bf 61}, 365 (1988).

\bibitem{LSCO-pi-0}
J. P. Hill, G. Blumberg, Young-June Kim, D. S. Ellis, S. Wakimoto,
R. J. Birgeneau, Seiki Komiya, Yoichi Ando, B. Liang, R. L. Greene,
D. Casa, and T. Gog, Phys. Rev. Lett. {\bf 100}, 097001 (2008). 

\bibitem{dd-93}
J. D. Perkins, J. M. Graybeal, M. A. Kastner, R. J. Birgeneau, J. 
P. Falck, and M. Greven, Phys. Rev. Lett. {\bf 71}, 1621 (1993).

\bibitem{dd-94}
J. P. Falck, J. D. Perkins, A. Levy, M. A. Kastner, J. M. Graybeal, 
and R. J. Birgeneau, Phys. Rev. B {\bf 49}, 6246 (1994). 

\end{thebibliography}
\end{document}